\documentclass[final,3p,times,twocolumn]{elsarticle}

\usepackage{bm}  
\usepackage{amsmath}
\usepackage{booktabs} 
\usepackage[normalem]{ulem}
\usepackage{subcaption}
\usepackage[font=small,skip=4pt]{caption}
\usepackage{ltablex}
\usepackage{enumitem}
\usepackage[nolist, nohyperlinks]{acronym}
\usepackage{multirow}
\usepackage{color}

\newcolumntype{L}[1]{>{\raggedright\let\newline\\\arraybackslash\hspace{0pt}}m{#1}}
\newcolumntype{C}[1]{>{\centering\let\newline\\\arraybackslash\hspace{0pt}}m{#1}}
\newcolumntype{R}[1]{>{\raggedleft\let\newline\\\arraybackslash\hspace{0pt}}m{#1}}

\journal{Graphical Models}

\begin{document}

\begin{frontmatter}



\title{3D Mesh Segmentation via Multi-branch 1D Convolutional Neural Networks}


\author{David George}
\author{Xianghua Xie}
\author{Gary KL Tam}

\address{Swansea University, United Kingdom}

\begin{abstract}
There is an increasing interest in applying deep learning to 3D mesh segmentation. We observe that 1) existing feature-based techniques are often slow or sensitive to feature resizing, 2) there are minimal comparative studies and 3) techniques often suffer from reproducibility issue. This study contributes in two ways. First, we propose a novel convolutional neural network (CNN) for mesh segmentation. It uses 1D data, filters and a multi-branch architecture for separate training of multi-scale features. Together with a novel way of computing conformal factor (CF), our technique clearly out-performs existing work. Secondly, we publicly provide implementations of several deep learning techniques, namely, neural networks (NNs), autoencoders (AEs) and CNNs, whose architectures are at least two layers deep. The significance of this study is that it proposes a robust form of CF, offers a novel and accurate CNN technique, and a comprehensive study of several deep learning techniques for baseline comparison.
\end{abstract}


\end{frontmatter}

\begin{acronym}
	\acro{ReLU}{Rectified Linear Unit}
	\acro{conv}{convolution}
	\acro{NN}{Neural Network}
	\acrodefplural{NN}[NNs]{Neural Networks}
	\acro{AE}{Autoencoder}
	\acrodefplural{AE}[AEs]{Autoencoders}
	\acro{RF}{Random Forest}
	\acrodefplural{RF}[RFs]{Random Forests}
	\acro{CNN}{Convolutional Neural Network}
	\acrodefplural{CNN}[CNNs]{Convolutional Neural Networks}
	
	\acro{GC}{Gaussian curvature}
	\acro{CF}{conformal factor}
	\acro{PC}{principal curvature}
	\acro{PCA}{principal component analysis}
	\acro{SDF}{shape diameter function}
	\acro{DMS}{distance from medial surface}
	\acro{AGD}{average geodesic distance}
	\acro{SC}{shape context}
	\acro{SI}{spin images}
	\acro{HKS}{heat kernel Signature}
	\acro{SIHKS}{scale invariant \ac{HKS}}

	\acro{PSB}{Princeton Segmentation Benchmark}
	
\end{acronym}

\section{Introduction}
\label{sec:intro}
Automatic mesh segmentation is the decomposition of a 3D mesh into meaningful parts. It aims to produce results as similar to those produced by humans. The ability to properly segment a 3D mesh is important to many downstream applications, such as shape retrieval \cite{shapira2010}, matching \cite{kleiman2015}, editing \cite{yu2004,chen2015a}, deformation \cite{yang2013} and modelling \cite{chen2015b}. Many of these applications require well-defined mesh segments, making a robust and accurate segmentation algorithm essential.

From a machine learning point of view, mesh segmentation can be broadly categorised as unsupervised and supervised segmentation.
Earlier techniques focused on segmenting a single mesh in an unsupervised manner. They first compute features (e.g. shape diameter function \cite{shapira2008}, approximate convexity \cite{kaick2014} and curvature \cite{meyer2002}) for the faces of the meshes, and uses an optimisation technique to produce the segmentation results. Notable techniques include k-means \cite{shlafman2002}, mean-shift clustering \cite{comaniciu2002} and normalized and randomized cuts \cite{golovinskiy2008}. A detailed survey can be found in \cite{shamir2008,chen2009}. Given the large shape variability of segments, more recent approaches consider consistent co-segmentation of a collection of shapes, where class labels are consistent throughout the set \cite{sidi2011,huang2011}.

Segmentation often requires a higher level understanding of the 3D shapes, as composition of an object often relates to  shapes and functionality of its parts  \cite{hu15icon} . Supervised techniques treat segmentation as a labelling problem and use machine learning to optimise the mapping from features to labels. It requires extensive manual effort to label all data properly for training. With the recent effort from the community (e.g. shape benchmarks \cite{chen2009}), supervised techniques are gaining focus. 
The work by \cite{kalogerakis2010} pioneered to apply joint boosting on a large set of shape features for effective labelling. Recently, \cite{xie2014} used an extreme learning machine (a single layer wide neural network), then later expanded it to a two-layer network \cite{xie2015}, however the performance is marginally better than traditional shallow classifiers \cite{evangelos2017}. Later \cite{guo2015} applied \acp{CNN} to mesh segmentation.

Despite these research efforts, there are still many research questions unexplored. First, it is generally unclear which deep learning techniques work and which do not for mesh segmentation, and what features work best. To the best of our knowledge, most supervised feature-based mesh segmentation techniques use geometric features derived mostly from one face (except [18] which also computes a few geometric features curvatures, PCAs with different radii). As such, spatial scale information is mostly not considered in their learning architecture. A feature-based deep learning \emph{network architecture} for segmentation that considers multi-scale geometric features derived from a set of local faces has not been fully explored. Also, there has been no comparative analysis of a broader spectrum of deep learning techniques. Second, the reproducibility of these techniques depends on the architecture, exact implementation and the set of training datasets used. This information and along with complete source code is largely unavailable. Coupled with these, there are also challenges in training the networks properly due to the variability in \ac{CNN} architectures, large number of samples (200K-2M samples per set) and lengthy training time (in terms of months). All these elements hinder the development of supervised 3D segmentation techniques.

In this paper, we try to address several research questions. (i) Compared to existing learning techniques that use features mostly defined per face, can a deep learning architecture, considering multi-scale features derived from a set of faces, be useful? (ii) Compared to \cite{guo2015} that reshapes features into 2D images and applies a basic image-based \ac{CNN} pipeline for shape segmentation, can we treat input features as a single 1D feature vector? This would avoid the tuning of image size, and improve efficiency and performance of \acp{CNN}. (iii) Finally, how much improvement can \acp{CNN} have over existing deep learning techniques. Our contributions of this paper are four-fold:

\begin{itemize}[leftmargin=*]
\setlength\itemsep{-2pt}
	\item First, we introduce a novel and accurate \ac{CNN} technique for 3D mesh segmentation. We introduce a multi-branch network architecture that separately trains features of three different scales. These multi-scale features are derived from features that associates to an increasing local neighbourhood of faces. The use of 1D feature vectors also remove most of the assumed feature relationships that are imposed by an image-based \ac{CNN} when reshaping the feature vector into a 2D image. Our novel technique clearly out-performs existing feature-based \ac{CNN} technique \cite{guo2015}.
	\item Second, we propose a novel feature vector of \ac{CF} which is computed from incremental smoothing of geometry. It is less sensitive to high curvature noise, and consistently provides higher segmentation accuracy than \cite{ben2008} alone.
	\item Third, we perform a comprehensive comparison of deep learning techniques (at least two layers deep) for supervised mesh segmentation, specifically \acfp{NN}, \acfp{AE} and \acp{CNN} \cite{guo2015}, showing the strengths and limitations of each technique by comparing their accuracies.
	\item Finally, data and our implementations of all the compared techniques are made publicly available for the research community.
\end{itemize}

The rest of the paper is structured as follows: Section~\ref{sec:relatedwork} provides a more detailed summary of both supervised and unsupervised mesh segmentation techniques. Section~\ref{sec:method} discusses different methods we compared, and our proposed technique using multi-branch 1D \ac{CNN} and multi-scale features for 3D segmentation. Section~\ref{sec:results} discussed our experiments and the results of all methods tested. Finally, Section~\ref{sec:conclusion} concludes this paper.

\begin{figure}
	\begin{center}
		\includegraphics[width=1.0\columnwidth]{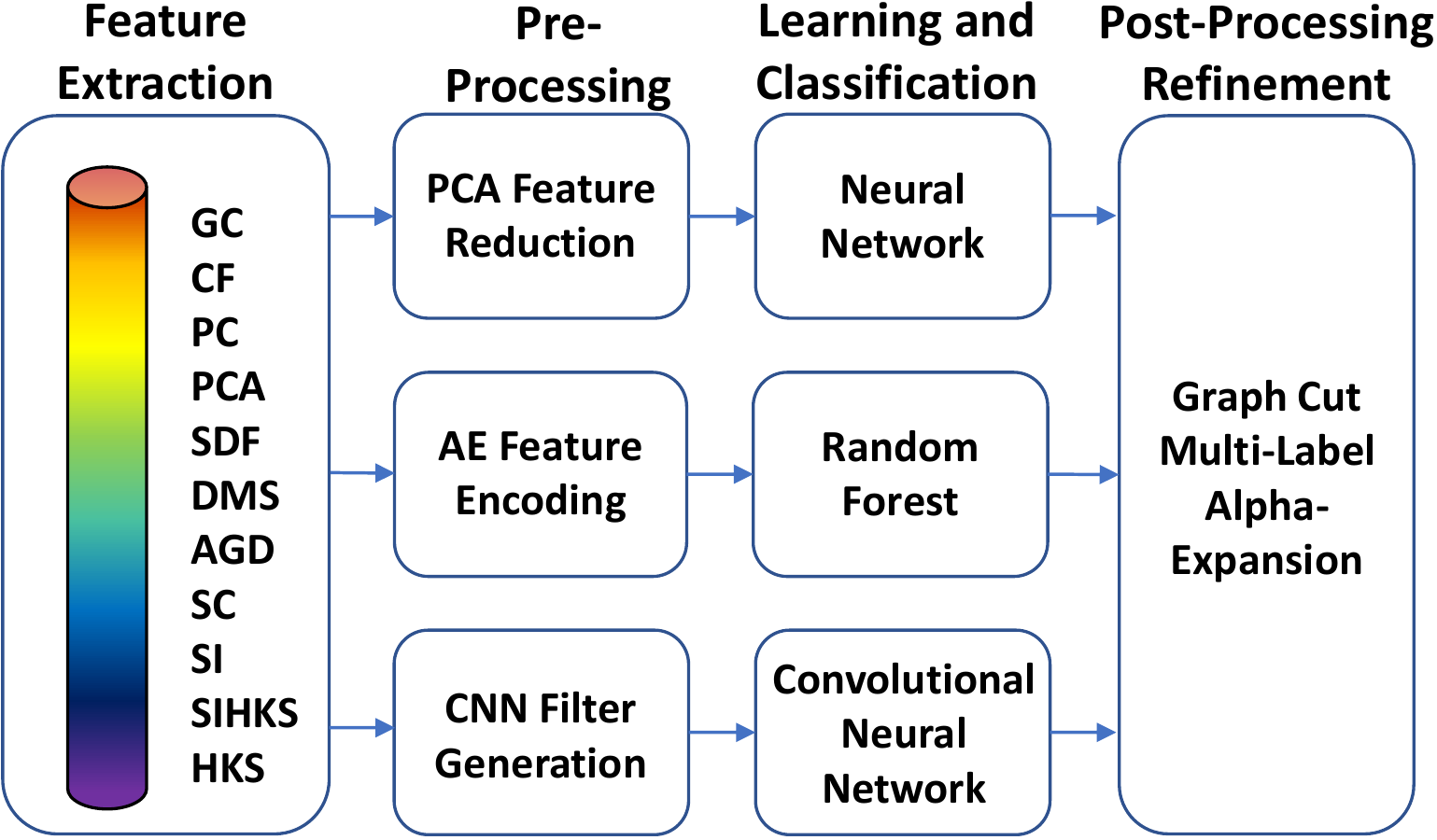}
	\end{center}
	\caption{An overview of the stages involved in each of the techniques. Each technique includes all four stages: feature extraction, pre-processing (by encoding or reduction), deep learning and classification, and graph-cut post-processing stage.}
	\label{fig:overview}
	\vspace{-0.5cm}
\end{figure}

\section{Related Work}
\label{sec:relatedwork}
This section first surveys existing techniques, with an emphasis on supervised segmentation. We then discuss the problems of existing supervised techniques, leading to our contributions.

\paragraph*{Unsupervised Segmentation}
Early work focused on simple, yet effective ideas for segmenting a single mesh \cite{shlafman2002,shapira2008}. They often performed clustering on geometric features (e.g. shape diameter function \cite{shapira2008}, geodesic distances \cite{hilaga2001}, curvature \cite{gal2006}) or partitioned based on properties that can be derived from the mesh itself (e.g. skeleton \cite{shapira2008}, convexity \cite{kaick2014}, fitting primitive shapes \cite{attene2006}). Many of these ideas have been shown effective, giving rise to a wide range of shape descriptors, and segmentation techniques, supporting many downstream applications \cite{shamir2008,chen2009}. However, segmenting a single mesh using a few features is often difficult due to the large variations in terms of shape and topology, even within the same class of objects. Recent research has adopted the co-analysis framework to investigate consistent segmentation of a collection of shapes from a single object class \cite{sidi2011,hu2012,meng2013,wu2013, shu2016}. For example, all legs in a chair set should be labelled the same. Such constraint is powerful yet requires less human effort. However, these methods rely on consistent geometric similarity within the set and a reliable shape/part matching algorithm in order to perform well. The large variations between different shapes in the same set and the sparse number of shapes in the set often cause problems in the final segmentation \cite{theologou2015}. More importantly, segments of a shape are often associated to its functionality - a high-level understanding of shapes \cite{hu15icon}. Therefore, there is an increasing interest in supervised segmentation techniques, trying to learn a high-level mapping directly from feature to segment.

\begin{figure}
	\begin{center}
		\includegraphics[width=0.77\columnwidth]{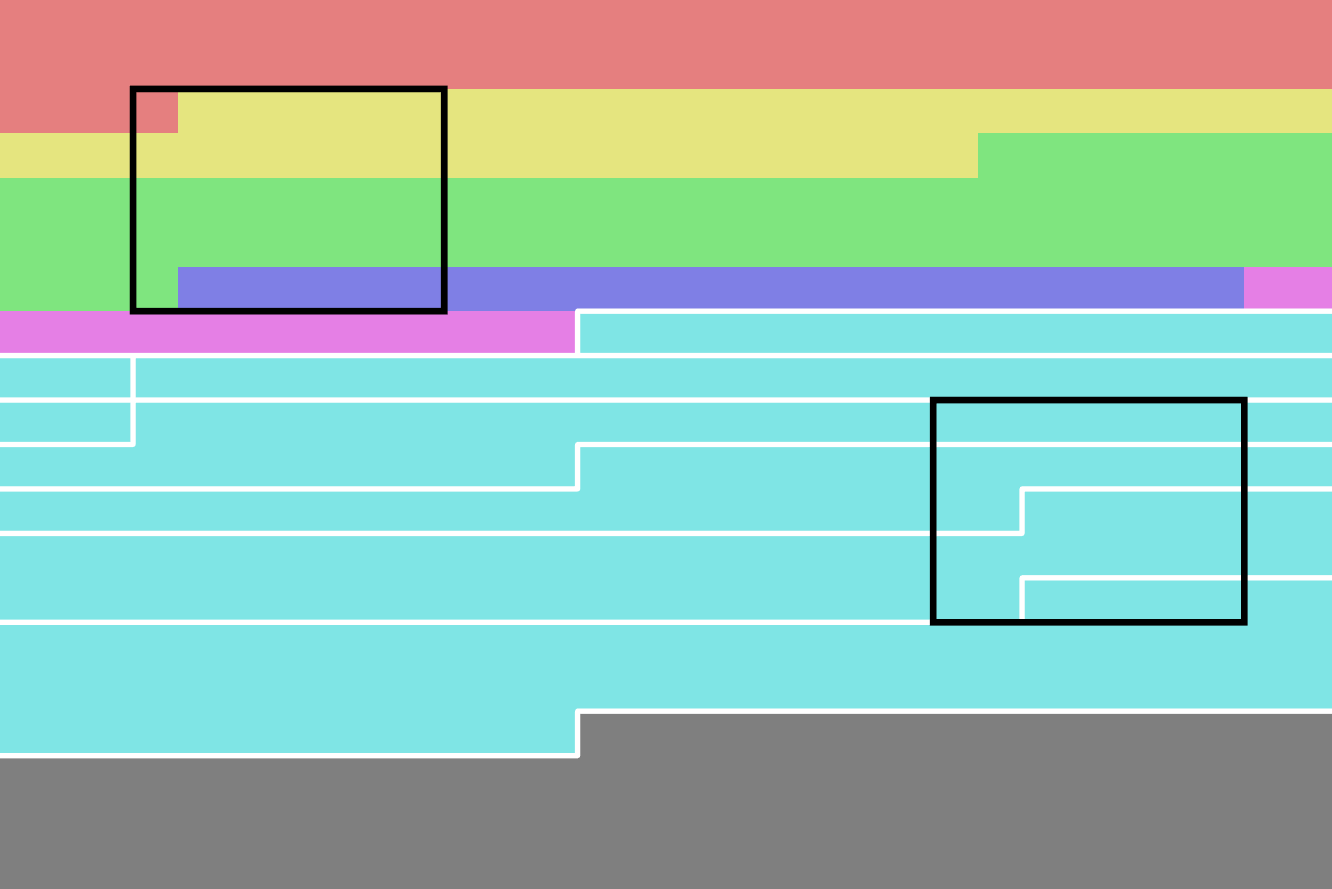}
	\end{center}
	\caption{Example of a 30x20 image produced by reshaping the 600 features in \cite{guo2015}. Each colour represents a distinct feature (\ac{PC}, \ac{PCA}, \ac{SDF}, \ac{DMS}, \ac{AGD}, \ac{SC}, \ac{SI} from top to bottom) and the white boarders outline the 6 different \ac{SC} histograms. The black rectangles are examples of 7x5 convolutional filters being passed over the image. In the examples, the filters would infer relationships between 4 different features or arbitrary bins in 4 different SC histograms, which in both cases have no correlation.}
	\label{fig:2dFeatures}
	\vspace{-0.5cm}
\end{figure}

\paragraph*{Supervised Segmentation}
These techniques treat 3D mesh segmentation as a labelling problem and use machine learning to optimise the mapping from features to labels. It requires extensive manual effort to label all data. The recent effort from the community contributed to a large set of segmentation benchmarks (e.g. \cite{chen2009}).

Existing supervised techniques rely on local features. The work by \cite{kalogerakis2010} proposed a method for mesh segmentation where a large pool of geometric features are ranked using JointBoost so that the best features are used to describe specific segments. Similarly, the work by \cite{benhabiles2011} ranks a large pool of features in order to detect the optimal segment boundaries for a given mesh, and an extreme learning machine was trained to classify labels using one \cite{xie2014} and two layers \cite{xie2015}.
However, supervised methods can perform poorly on very complex meshes, due to insufficient training data or large variations within label classes \cite{xie2014,guo2015}.

Recently, \cite{guo2015} extended the \ac{CNN} idea to 3D segmentation. They reshape a large pool of geometric features into a matrix resembling that of a 2D image, fitting the 2D image-based \ac{CNN} pipeline, and then train a \ac{CNN} on these ``images'' using the ground truth labels. The technique shows good performance, however, the reshaping and the use of 2D filters may infer relationships between adjacent rows of features that may have no correlation. As Figure~\ref{fig:2dFeatures} shows, passing a convolutional filter over such an ``image'' would unavoidably infer relationships between (up to 5) unrelated features regardless of the position of the filter.

From the literature, we have two further observations. First, existing feature driven techniques use local features developed by the influential work \cite{kalogerakis2010}. These features are mostly defined per face (a few are normalized by different geodesic radii for smoothing purposes), as such, there is no spatial scale information included in the architecture. To the best of our knowledge, a feature-based deep learning network architecture that considers multi-scale geometric features derived from a set of local faces has not been fully explored in supervised mesh segmentation. We hypothesise that multi-scale features would be useful because face-based \cite{sidi2011} and patch-based techniques \cite{wu2013} have both shown good performance in co-segmentation. Second, whilst deep learning is useful, there is not much analysis as how \acp{CNN} perform compared to other techniques in the deep learning family.

\begin{figure}
	{\centering
		\begin{tabular}{@{}c@{}c@{}c@{}c@{}}
			\includegraphics[width=0.242\columnwidth]{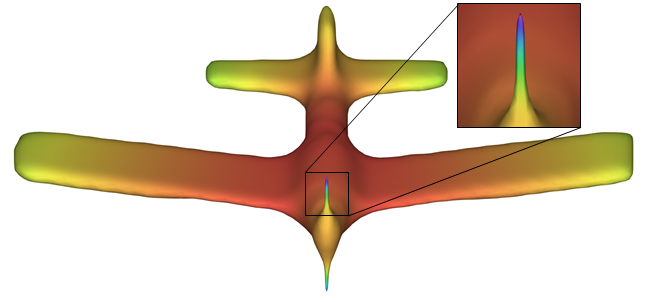}&
			\includegraphics[width=0.242\columnwidth]{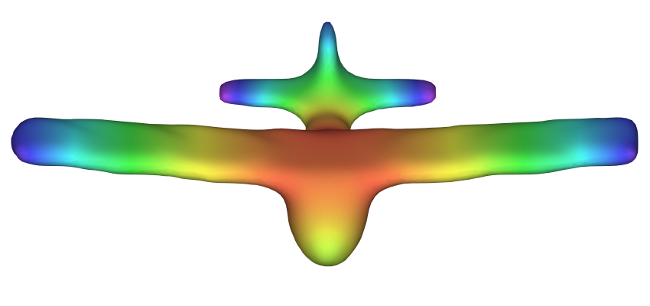}&
			\includegraphics[width=0.242\columnwidth]{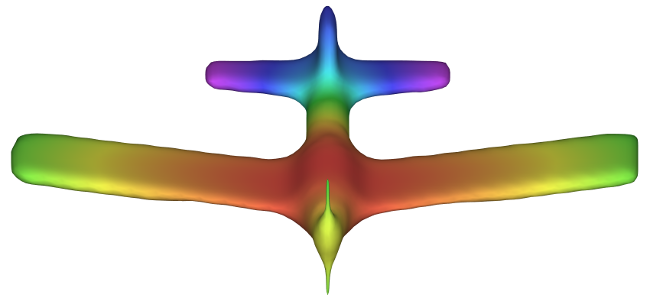}&
			\includegraphics[width=0.242\columnwidth]{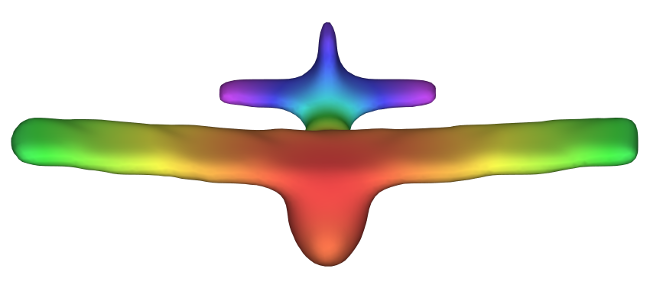} \\
			\includegraphics[width=0.2\columnwidth]{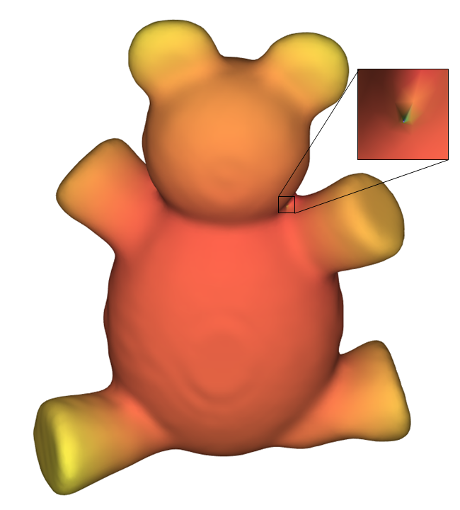}&
			\includegraphics[width=0.2\columnwidth]{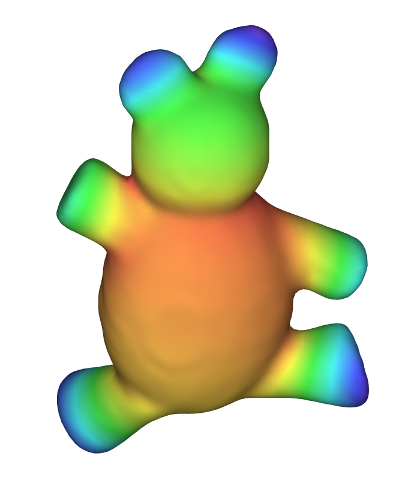}&
			\includegraphics[width=0.2\columnwidth]{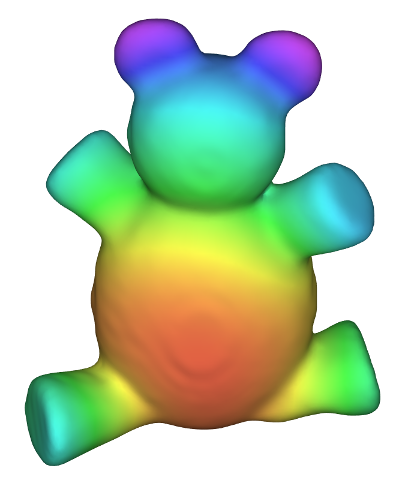}&
			\includegraphics[width=0.2\columnwidth]{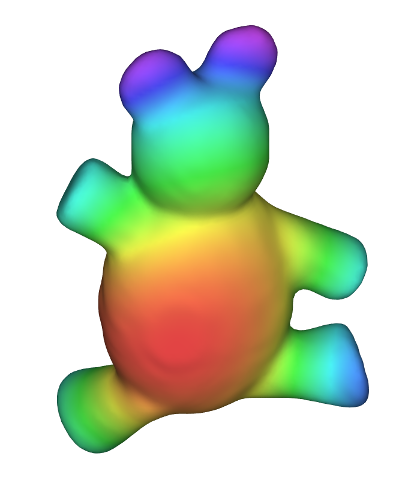} \\
			\begin{subfigure}{0.242\columnwidth}
				\includegraphics[width=\columnwidth]{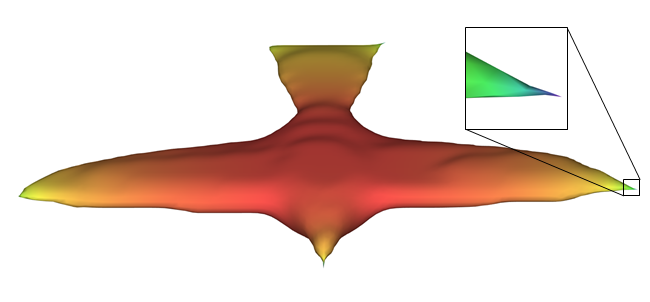}
				\caption{}
			\end{subfigure} &
			\begin{subfigure}{0.242\columnwidth}
				\includegraphics[width=\columnwidth]{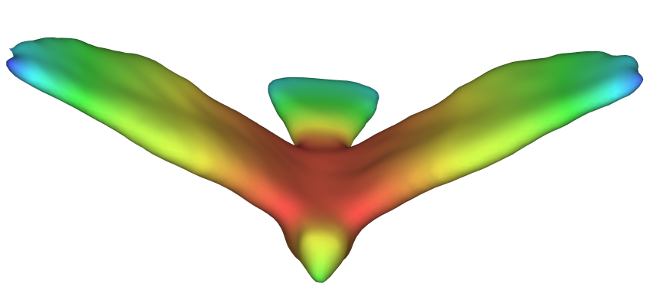}
				\caption{}
			\end{subfigure} &
			\begin{subfigure}{0.242\columnwidth}
				\includegraphics[width=\columnwidth]{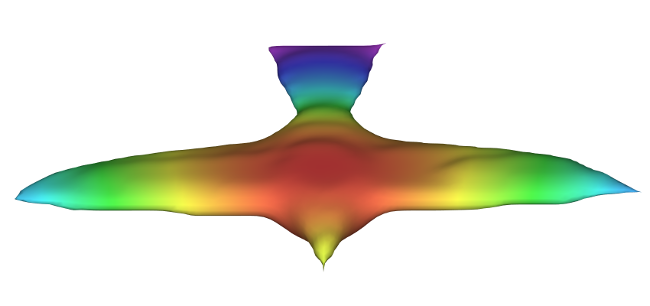}
				\caption{}
			\end{subfigure} &
			\begin{subfigure}{0.242\columnwidth}
				\includegraphics[width=\linewidth]{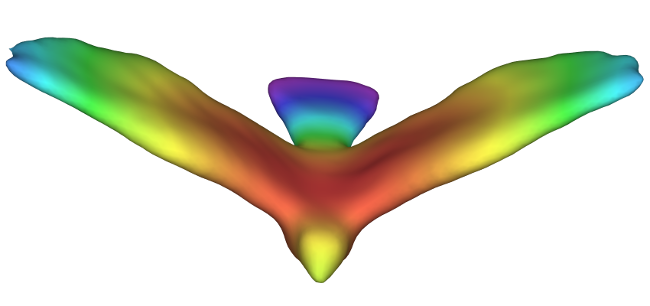}
				\caption{}
			\end{subfigure}
		\end{tabular}\par}
	\caption{Visual comparison of \acf{CF} features. Columns (a) and (b) are the original \ac{CF}, columns (c) and (d) are \ac{CF} after one stage of Laplacian-smoothing. \ac{CF} is sensitive and can be easily distorted by small regions of large curvature (propeller tips of the plane, noise on shoulder of teddy and wing tips of the bird in column (a)). Non-shrinking Laplacian-smoothing can alleviate the geometry issues, making the computed \ac{CF} much more consistent across similar meshes (columns (c) and (d))}
	\label{fig:cf}
\end{figure}

\begin{figure*}[t]
	\begin{center}
		\includegraphics[width=1.0\textwidth]{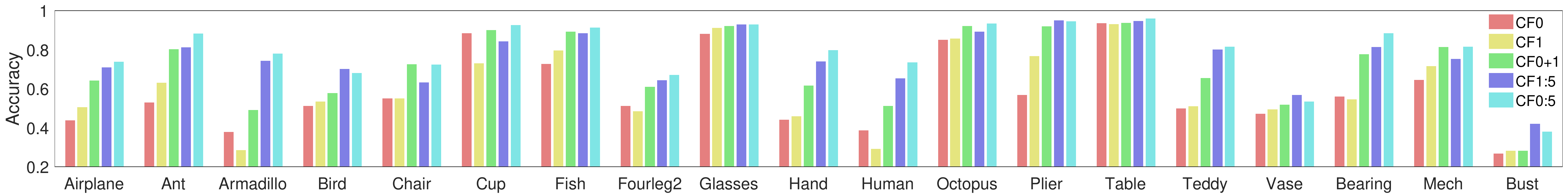}
	\end{center}
	\caption{Accuracies of running leave-one-out cross validation on all sets using Random Forest classifier. CF0, CF1, CF0+1, CF1:5 and CF0:5 respectively denote the original \ac{CF} \cite{ben2008}, \ac{CF} after one stage of Laplacian-smoothing, combination of CF0 and CF1, combination of all smoothed \ac{CF} features, and combination of all \ac{CF} (including CF0). The chart shows that the new smoothed \ac{CF} features improve upon the original in most cases, and also further improves when they are combined with the original \ac{CF}.}
	\label{fig:cfResults}
\end{figure*}
In this paper, we show that by using multi-scale features, treating them as separate 1D vectors per scale, and applying multi-branch 1D \ac{CNN} filters through the network, we avoid the parameter tuning problem of reshaping a 2D matrix. Our method clearly out-performs existing work \cite{guo2015} in accuracy. Further, we provide comprehensive evaluation of various deep learning techniques, and show how \acp{CNN}, though more complex, can improve performance over simpler architectures (\acp{NN}, \acp{AE}). It is worth noting that we have found existing methods lack reproducibility. Some existing work have not provided any or complete experimental code. Despite using the exact architecture and settings stated, some high performing results are hard to reproduce.

Concurrent to our work, there are recent efforts focusing on different kind of inputs, such as point clouds \cite{qi2016, qi2017}, octrees \cite{wang2017}, multiple projected images \cite{evangelos2017}, graphs \cite{yi2016} or geometry images \cite{maron2017}. 
Differing from these, our study focuses on feature-based approach, and investigates how deep learning can improve 3D segmentation using insightful geometry features that are developed in the past decade.


\section{Methodology}
\label{sec:method}
This section discusses the deep learning techniques proposed and evaluated. Section~\ref{sec:features} discusses geometric features used. We then summarise several techniques, Fully Connected \acfp{NN}, \acfp{AE} + \acfp{RF}, and \acfp{CNN}, in Sections~\ref{sec:neuralnet}-\ref{sec:cnn}, focusing on models which are at least two layers deep. Each technique is broken down into stages, namely, feature extraction, pre-processing, learning and classification, and post-processing (Figure~\ref{fig:overview}). Section~\ref{sec:refinement} describes the use of graph-cut \cite{boykov2001} for final refinement.

\subsection{Feature Extraction}
\label{sec:features}
To obtain a good feature representation of the meshes, we compute 11 types of geometric features, namely, the \acf{GC} \cite{meyer2002}, \acf{CF} \cite{ben2008}, \acf{PC} \cite{gal2006}, \acf{PCA} of local face centres \cite{kalogerakis2010}, \acf{SDF} \cite{shapira2008}, \acf{DMS} \cite{liu2009}, \acf{AGD} \cite{hilaga2001}, \acf{SC} \cite{belongie2002}, \acf{SI} \cite{johnson1999}, \acf{HKS} \cite{sun2009}, and \acf{SIHKS} \cite{bronstein2010}. They are calculated with different scales and normalisations. Most of these have been shown useful in earlier studies \cite{kalogerakis2010,guo2015}.

\ac{HKS} and \ac{SIHKS} have not yet been used in supervised mesh segmentation. They are effective point descriptors, designed for shape retrieval and correspondence \cite{bronstein2010}. As they are shown consistent in similar local regions, we hypothesise that they may be useful and include them in the feature set.

\ac{CF} has been used in unsupervised techniques (e.g. \cite{wu2013}) and been shown to be highly useful, but has not been used in supervised segmentation. When \ac{CF} is computed on meshes in small regions with large curvatures (e.g. the propeller of the left plane or wing tip of the bird in Figure~\ref{fig:cf}), \ac{CF} is seriously distorted. To resolve this issue, we introduce a multi-resolution version of \ac{CF}. We generate meshes with increasing number of smoothing iterations, using non-shrinking Laplacian smoothing \cite{taubin1995}, and compute \ac{CF} on these meshes. These new \acp{CF} alleviate the distortion and are more consistent (Figure~\ref{fig:cf}).

Let $M = \{F, V\}$ be a mesh, where $F$ and $V$ are the faces and vertices of the mesh. We first compute 5 iterations of non-shrinking Laplacian smoothing, where the input of the next iteration is the output of the previous. This gives us $M^s_i = \{F^s_i, V^s_i\}$ for iterations $i = 1,\dots,5$. To compute the \acf{CF} ($\Phi$) on the unsmoothed meshes we follow \cite{ben2008}, by solving:
\[L\Phi = K^{T} - K^{orig}\]
where $L$ is the Laplace-Beltrami operator, $K^{orig}$ is the \ac{GC} of the mesh \cite{meyer2002} and $K^{T}$ is the target \ac{GC}, which is the uniform curvature given by:
\[
k_{v}^T = \left(\sum\limits_{j \in V} \kappa_j \right) \frac{\sum\limits_{f \in F_v} \frac{1}{3} area(f)}
{\sum\limits_{f \in F} area(f)}
\]
where $k_{v}^T$ is the target \ac{GC} of vertex $v$, $\kappa_j$ is the $j$\textsuperscript{th} element of $K^{orig}$, $F_v$ the set of faces that share vertex $v$ and $area(f)$ the surface area of face $f \in F$. With a smoothed mesh $M^s_i$, the smoothed \ac{CF} $\Phi^{s}_i$ is computed by solving:
\[
L\Phi^s_i = K^{sT}_i - K^{T}
\]
where $\Phi^{s}_i$ is the desired \ac{CF} and $K^{sT}_i$ is the target \ac{GC} for the smoothed mesh $M^s_i$, and $K^T$ is the target \ac{GC} of the original mesh $M$. 
The rationale of using $K^T$ in our formula (instead of $K^{orig}$ of the smoothed mesh $M^s_i$) is that we would like the new \ac{CF} to model the changes due to geometry smoothing alone. We do not want it to be affected by the underlying tessellation as in the original \ac{CF} formula, making our \ac{CF} more robust.

To show the impact of the proposed \ac{CF} features we ran several experiments on the \ac{PSB} \cite{chen2009}. Each experiment used one or many of the \ac{CF} features to train a \ac{RF} classifier for mesh segmentation. Leave-one-out cross validation was performed on each set, with 3 replicates ran per tested mesh. The results for each experiment for each set is shown in Figure~\ref{fig:cfResults}. This shows that, in the majority of cases, the proposed \ac{CF} features have a large positive impact on the performance for classification, and also in some cases, just using a single smoothed \ac{CF} feature is better than using the original \ac{CF} feature. 

In total, we obtain an 800-component feature vector for each face. The vector consists of 593 features from \cite{kalogerakis2010}, 1 original + 5 new \ac{CF} features, 1 GC feature, 100 HKS features and 100 SIHKS features. They are used in all our techniques.

\begin{figure*}[t]
	\begin{center}
		\includegraphics[width=1.0\textwidth]{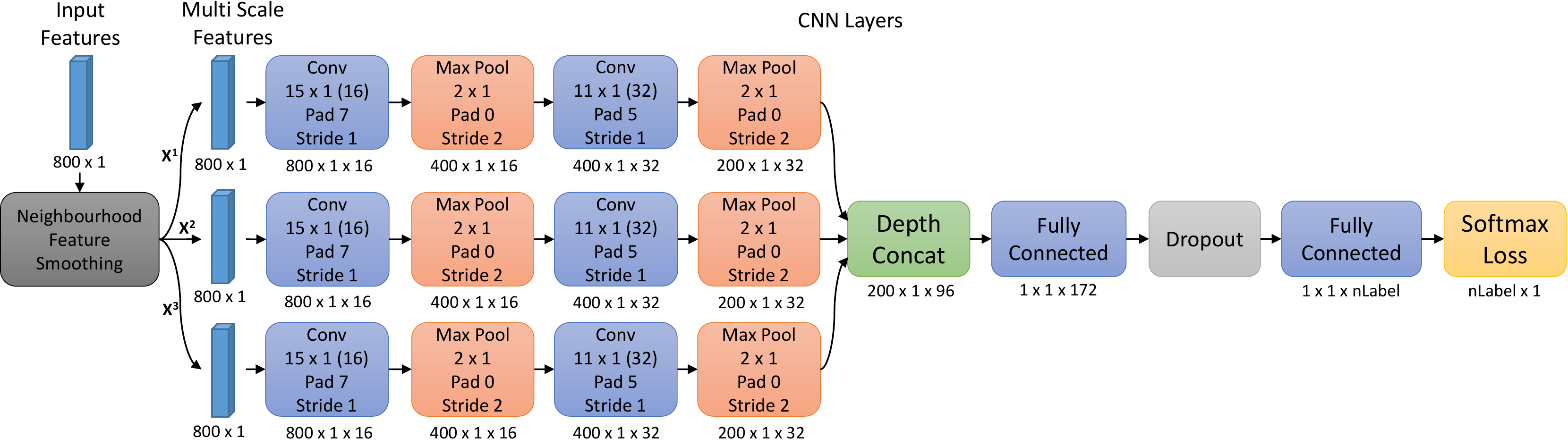}
	\end{center}
	\caption{The architecture of our multi-scale 1D \ac{CNN}. Given an 800-dimension feature vector $\bm{X}^1$ of a face $u$, we compute a set of surrounding faces $\mathcal{N}^k(u)$ that are $k-1$ steps away ($k = 2,3$). We average all features of all faces in $\mathcal{N}^k(u)$ leading two extra feature vectors $\bm{X}^{2,3}$. These multi-scale features $\bm{X}^{1...3}$ are used in the \ac{CNN}, and trained separately through the network. They are then concatenated by the depth concatenation layer before reaching the fully connected and classification layers. Each \ac{conv} layer is followed by a batch normalization and a leaky \ac{ReLU} layer, and the first fully connected layer is followed by a leaky \ac{ReLU} layer.}
	\label{fig:architecture}
\end{figure*}

\subsection{Fully Connected Neural Network}
\label{sec:neuralnet}
The first deep learning technique we analysed for mesh segmentation was conventional Fully Connected \ac{NN}. \acp{NN} consist of several fully connected layers followed by a classification layer to produce prediction probabilities. Each layer consists of a set of neurons, each with a weight and a bias. Each feature is passed to every neuron, which is activated based on the output of an activation function.
The output from each neuron is then fed to every neuron in the next layer.

\acp{NN} iterates between two stages, a feed-forward and a back-propagation pass. The feed-forward pass is an unsupervised stage where all input feature is passed through the network. The neurons and their internal parameters across all layers determine the response. The back-propagation pass is supervised and minimizes the errors between the ground truth and the predicted labels. Error values are passed back through the network in order to tweak all the internal parameters. This process is repeated (typically several hundred times) in order to fine tune all parameters such that the network best describes the mapping between features and labels \cite{liu2016}.

We perform \ac{PCA} to select the most important 50 principal components as features. This significantly reduces training time but does not affect the accuracy empirically. The reduced features are used to train a three-layer \ac{NN}. The first layer has the same number of neurons as input features (50 neurons), and each subsequent layer reduces the number of neurons by half (25 and 12 neurons).
The third layer is then fed into a softmax layer to compute an error cost. This then propagates back through the network to optimise the parameters. Once the network is trained, a new mesh is fed through for testing. The network returns a set of probabilities, specifying a class that a face may belong to, and allowing further refinement via graph-cut (Section~\ref{sec:refinement}). Results are reported in Table \ref{table:results}, \textbf{\ac{PCA} \& \ac{NN}} column.

\subsection{Autoencoder and Random Forest}
\label{sec:randomforest}

Next, we analyse the use of \acp{AE} for feature reduction and a \acp{RF} for classification. An \ac{AE} is a type of artificial \ac{NN}, which aims to encode features for dimensionality reduction. The idea of \ac{AE} is to learn the optimal representation of the original features through a network by recovering the original data through encoding and decoding \cite{liu2016}. It is powerful for its ability to non-linear dimension reduction. \acp{AE} can be stacked and optimised, so that the encoded features from one \ac{AE} can be fed to another for further reduction. Once trained, the encoded features are used to train a classifier.

In our technique, we pre-trained two \ac{AE} layers separately. The first layer takes 800 features and reduces it to 400. The second layer takes the 400 encoded features and reduces it to 200. These encoded features are then used to train a softmax layer. Once it is trained, all three (two \acp{AE} and a softmax) layers are stacked together and re-trained. The model can then be used for testing. Results are reported in Table \ref{table:results}, \textbf{AE \& NN} column.

The encoded features from the stacked \acp{AE} can also be used to train a \ac{RF} classifier. A \ac{RF} classifier is a learning technique that takes a large set of random decision trees (we used 100 trees) and averages their prediction. It offers high performance in accuracy and speed whilst avoids overfitting. Results are reported in Table \ref{table:results}, \textbf{AE \& RF} column. For both \textbf{AE \& NN} and \textbf{AE \& RF}, a graph-cut post-refinement (Section~\ref{sec:refinement}) is applied.

\subsection{Multi-scale 1D Convolutional Neural Network}
\label{sec:cnn}
Here we discuss our new \ac{CNN} mesh segmentation technique. We first outline the proposed multi-scale features, then describe the types of layers we use, and finally the \ac{CNN} architecture (Figure~\ref{fig:architecture}).

\paragraph*{Multi-scale feature extraction}
Existing techniques extract features mostly per face \cite{kalogerakis2010}. It has been shown in co-segmentation and relevant studies \cite{huang2011,wu2013} that patch would also be useful for segmentation. We thus hypothesise that multi-scale features derived from a set of neighbouring faces would be useful, and should be considered in a network architecture. Given a face $u$, we define a set of surrounding faces $\mathcal{N}^k(u)$ of $u$ as the surrounding faces that are at most $k-1$ step away. We then compute two extra feature vectors $\bm{X}^k$ where $k = 2,3$ by averaging feature values of all faces in $\mathcal{N}^k(u)$. This leads to three feature vectors $\bm{X}^k$ where $k=1...3$ and $\bm{X}^1$ is the original feature (Section \ref{sec:features}). Each of these feature vectors $\bm{X}^k$ are trained separately by the proposed \ac{CNN} network, and then merged before classification.

\paragraph*{Network layers}
We now discussed all network layers used in our \ac{CNN} architecture, and their functions.
\begin{itemize}[leftmargin=*]
	\setlength\itemsep{-2pt}
	\item \textbf{Convolutional layers} simulate the organisation of humans' visual cortex, and the neurons responses of local receptive field. They consist of filters, which are convolved over the input to produce new feature maps as outputs, one for each filter.
	\item \textbf{Batch normalization layers} typically follow \ac{conv} layers to normalise the output. It allows much higher learning rates, and makes the network less sensitive to the initialisation \cite{ioffe2015}.
	\item \textbf{Leaky ReLU layers} A \acf{ReLU} layer simulates the firing of a neuron by means of an activation function. We use the leaky \ac{ReLU} variation \cite{xu15icmlworkshop} instead of a regular \ac{ReLU} as having a small negative gradient will stop cases where all inputs are negative and the activation produces zero. We set the slope to be $0.2$ in our experiments.
	\item \textbf{Max pooling layers} Pooling layers are used to down-sample the output features of the previous layer to better manage the high feature size, these typically performs after the \ac{conv} layers.
	\item \textbf{Depth concatenation layers} To merge feature maps from different branches into a single feature map, we use a concatenation layer to concatenate via the depth dimension. This is pioneered in \cite{szegedy2015} to provide a mechanism for separate learning of features and later merging for classification.
	\item \textbf{Fully connected layers} (like \acp{NN}) have full connections to all activations in the previous layers, and act as the function approximators to learn the non-linear mapping.
	\item \textbf{Dropout layers} are included in to regularise the network to reduce overfitting \cite{srivastava2014}. It works by randomly selecting neurons to be ignored during this pass of the training. This is done by ignoring the weights assigned to them during the forward pass and not updating their weights on the back pass. We set 50\% of neurons to be randomly ignored in our experiments.
	\item \textbf{Softmax layers} are activation functions typically used for classification. The function computes the exponential of the input and divide them by the sum of all exponential values, giving prediction probabilities as output. Coupled with a loss function, they penalise differences between the predicted and true labels, and update the network parameters in back propagation.
\end{itemize}

\paragraph*{Architecture \& Rationale}
The architecture of our \ac{CNN} (Figure~\ref{fig:architecture}) allows the three multi-scale feature vectors to be trained independently before being merged back for the fully connected layers and classification. Each feature vector undergoes the same training process, with their own distinct layers and parameters. In our implementation, each \ac{conv} layer is followed by a batch normalisation and a leaky \ac{ReLU} layer. For clarity, these two layers are not shown in Figure \ref{fig:architecture}.

The rationale behind our architecture design stems from several research questions:
\begin{itemize}[leftmargin=0.6cm]
	\vspace{-5pt}
	\setlength\itemsep{-3pt}
	\item[Q1] Can we reduce the unnecessary inference between unrelated features and improve performance?
	\item[Q2] How can we make good use of multi-scale features in a deep learning architecture?
	\item[Q3] How can we train such a network in a practical time frame given the increased features and branches?
\end{itemize}

One possible answer to eliminate inference between unrelated features (Q1) is a fully connected \ac{NN}. However, such a network would lead to impractical training times (due to the increased feature sizes and number of scales). Our compromise is to use a 1D network instead. Though it does not fully resolve the issue in \cite{guo2015}, it avoids guessing the parameter for image resizing, and alleviates unnecessary inference between number of unrelated features from (at most) 5 to 2 per each filter. Further, because both face-level \cite{kalogerakis2010,guo2015} and patch-level features \cite{meng2013,shu2016} have been shown useful alone, we hypothesise that features from different scales can be trained and analysed independently (Q2). Inspired by GoogLeNet \cite{szegedy2015}, we separate and train features of each scale in individual branch, formulate our architecture as a multi-branch network and concatenate them through the depth channel. This also reduces training time (Q3). Fully connected layers are used to get the final predictions. We introduce a second fully connected layer due to the increased amount of data from all 3 branches. Finally the addition of batch normalization layers and a dropout layer allows the network to better generalize and reduce overfitting. We opted to use 3 branches as it shows highest accuracy with the quickest training time empirically (Q3). An evaluation on the use of 1-4 branches can be found in Section~\ref{sec:results}.

\begin{figure*}
	\begin{center}
		\begin{subfigure}[b]{0.245\linewidth}
			\includegraphics[width=\textwidth]{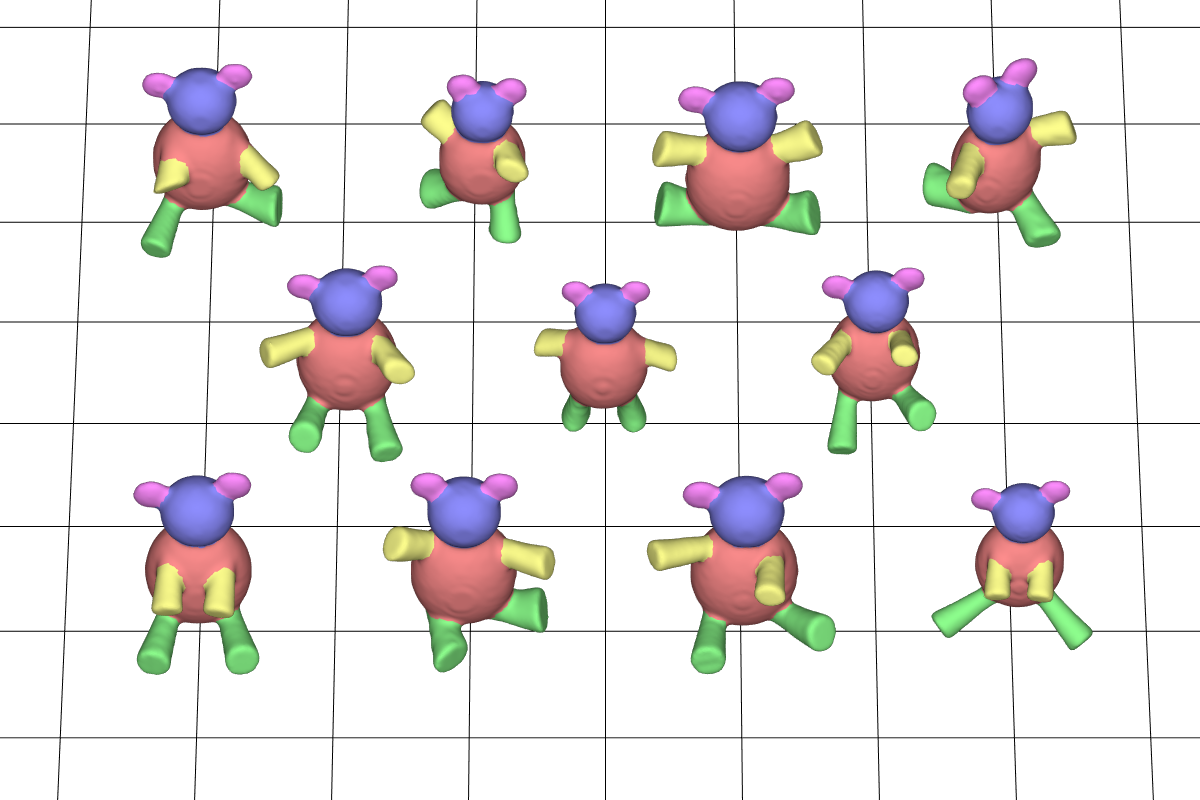}
			\caption{psbTeddy}
		\end{subfigure}
		\begin{subfigure}[b]{0.245\linewidth}
			\includegraphics[width=\textwidth]{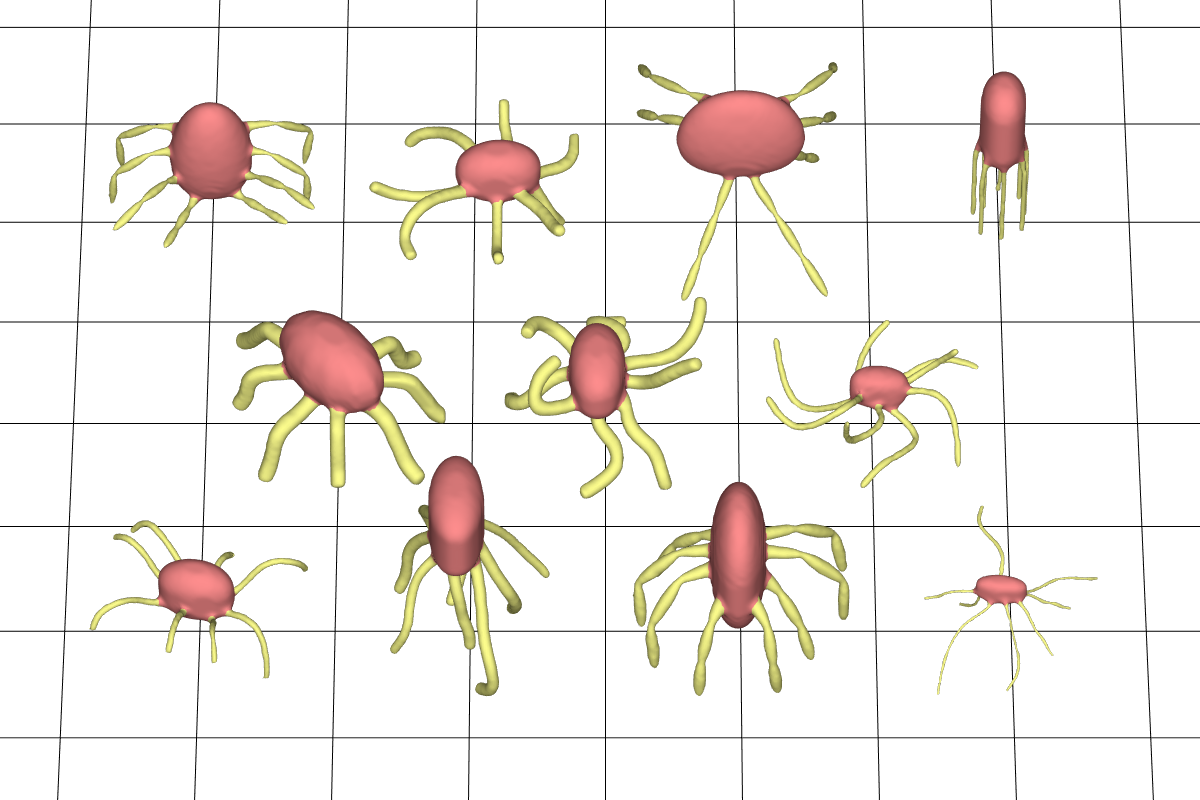}
			\caption{psbOctopus}
		\end{subfigure}
		\begin{subfigure}[b]{0.245\linewidth}
			\includegraphics[width=\textwidth]{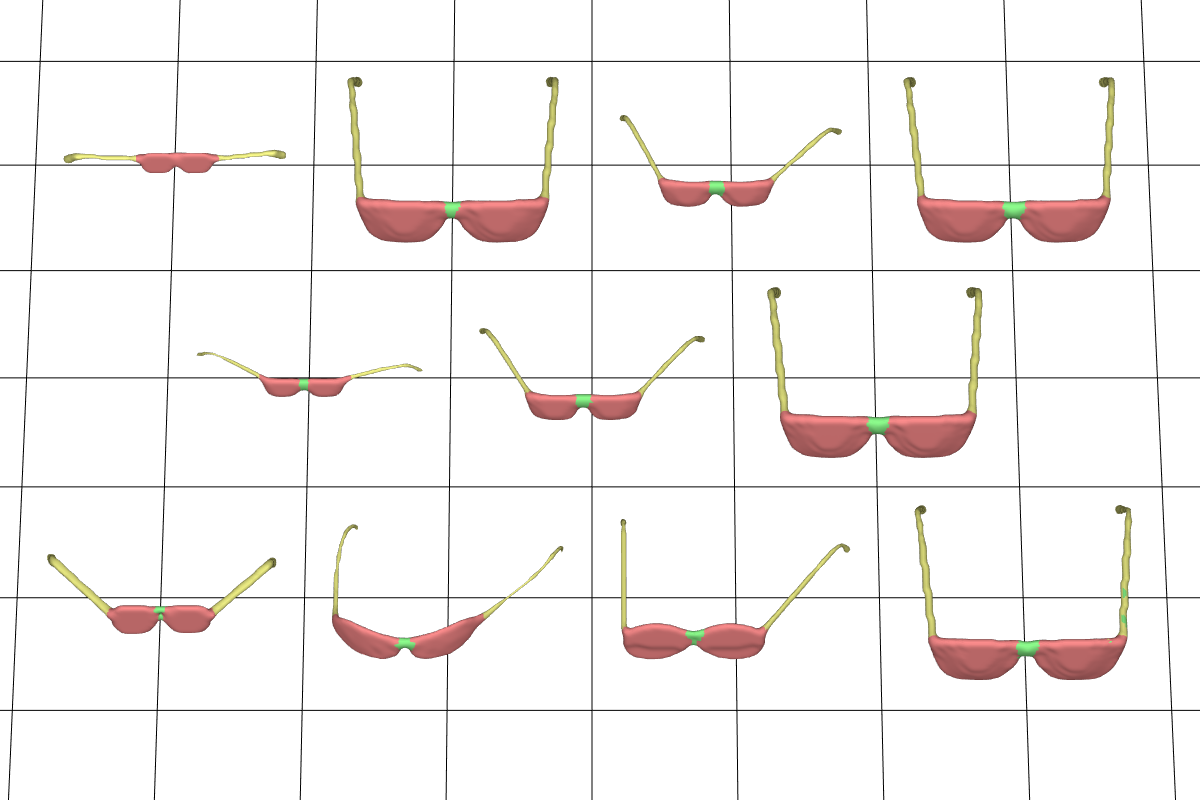}
			\caption{psbGlasses}
		\end{subfigure}
		\begin{subfigure}[b]{0.245\linewidth}
			\includegraphics[width=\textwidth]{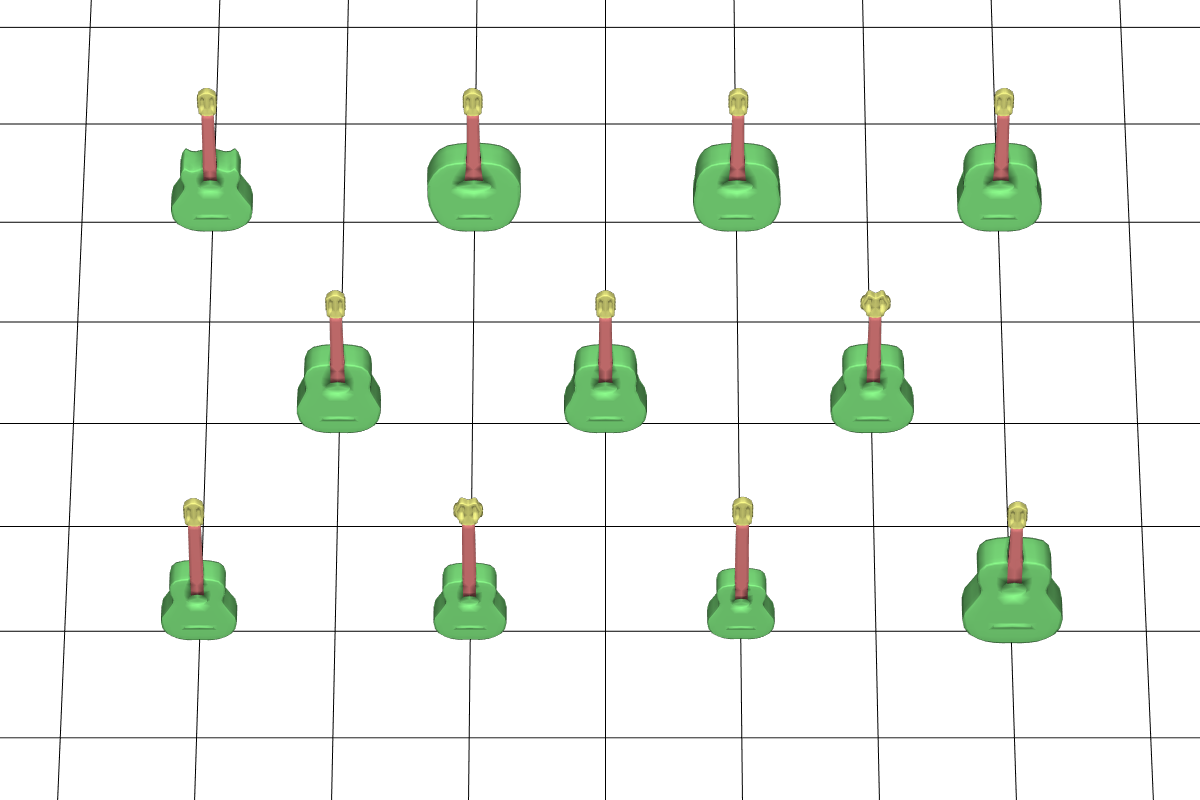}
			\caption{cosegGuitars}
		\end{subfigure}
		
		\begin{subfigure}[b]{0.245\linewidth}
			\includegraphics[width=\textwidth]{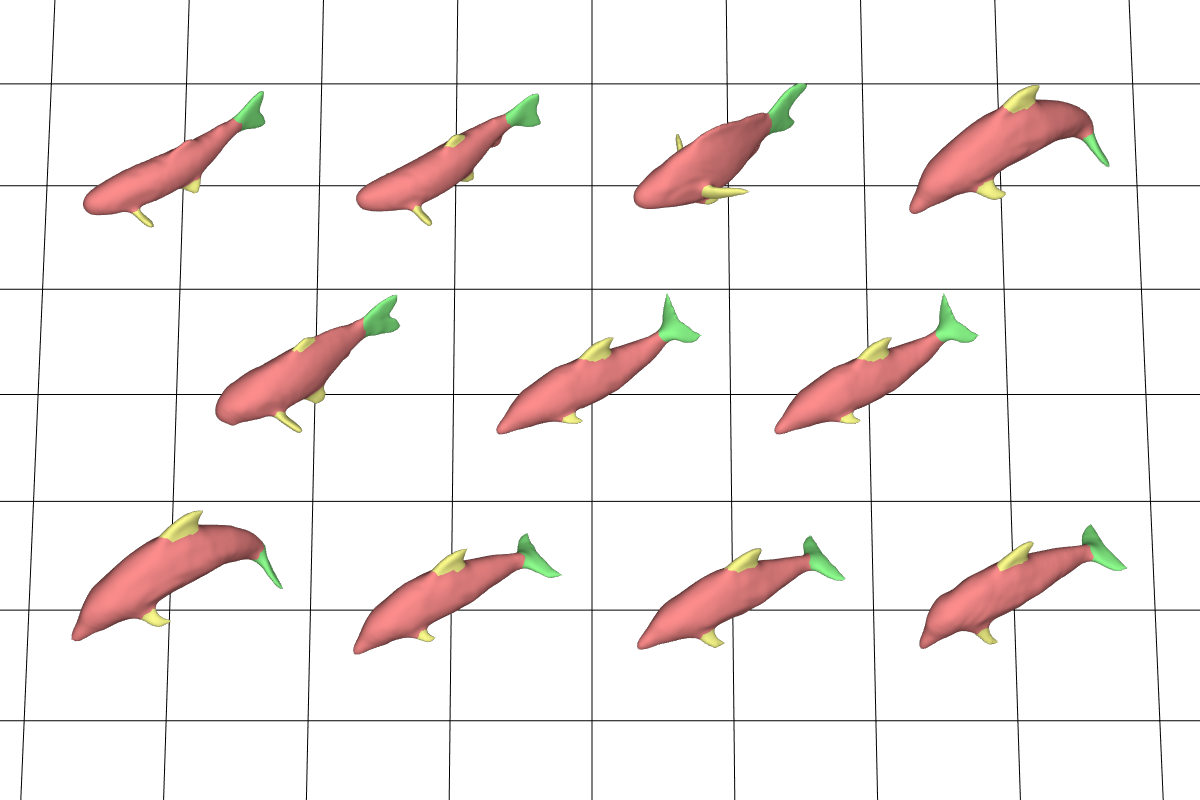}
			\caption{psbFish}
		\end{subfigure}
		\begin{subfigure}[b]{0.245\linewidth}
			\includegraphics[width=\textwidth]{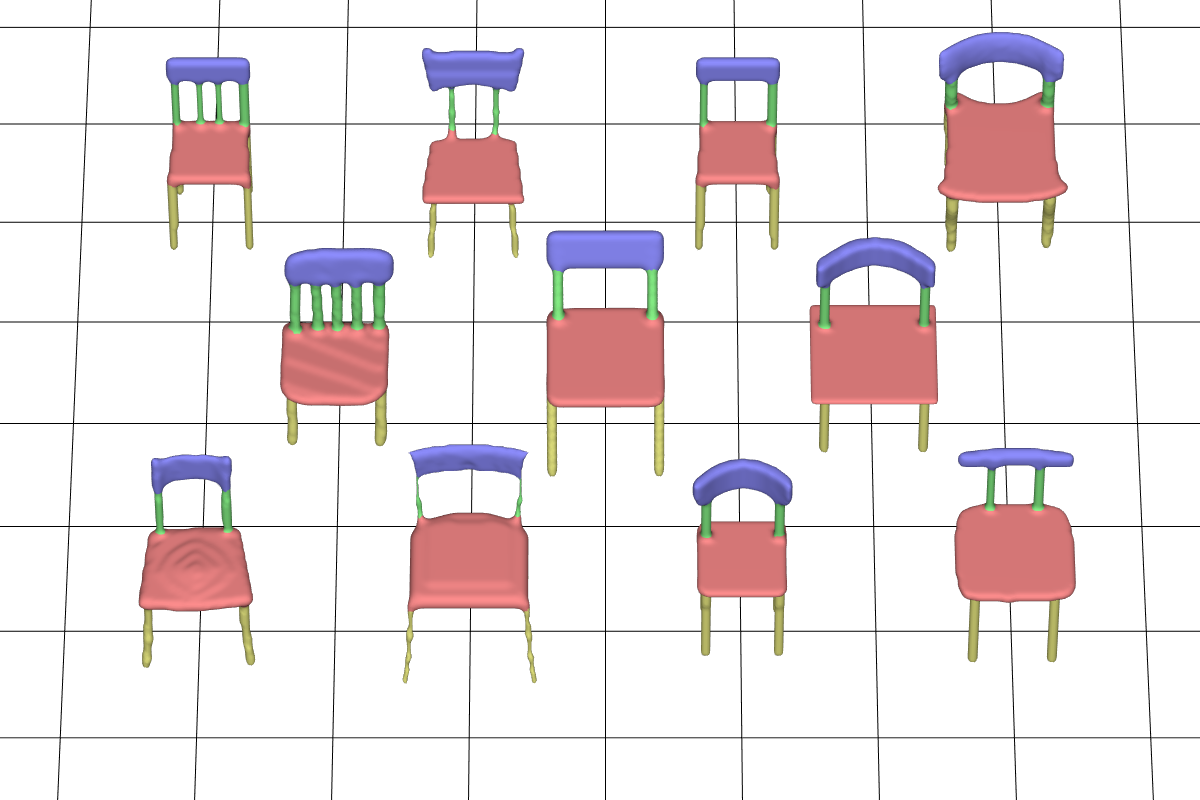}
			\caption{psbChair}
		\end{subfigure}
		\begin{subfigure}[b]{0.245\linewidth}
			\includegraphics[width=\textwidth]{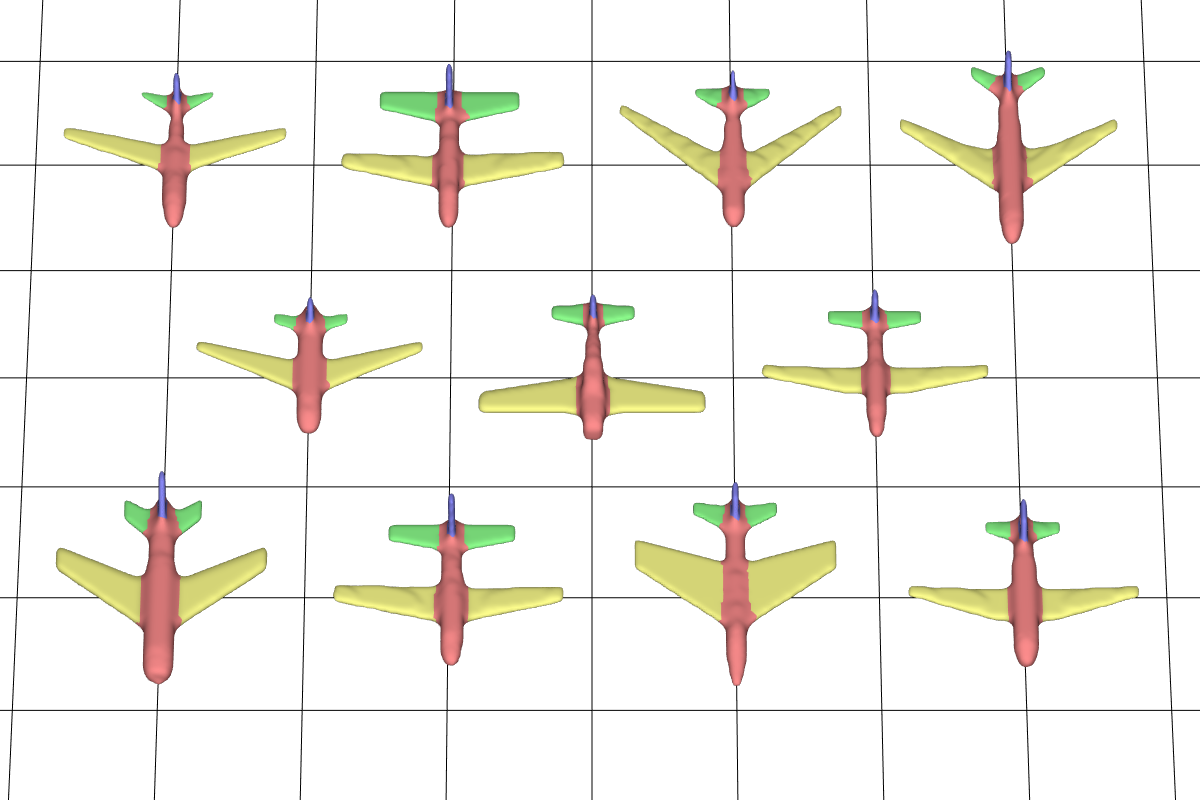}
			\caption{psbAirplane}
		\end{subfigure}
		\begin{subfigure}[b]{0.245\linewidth}
			\includegraphics[width=\textwidth]{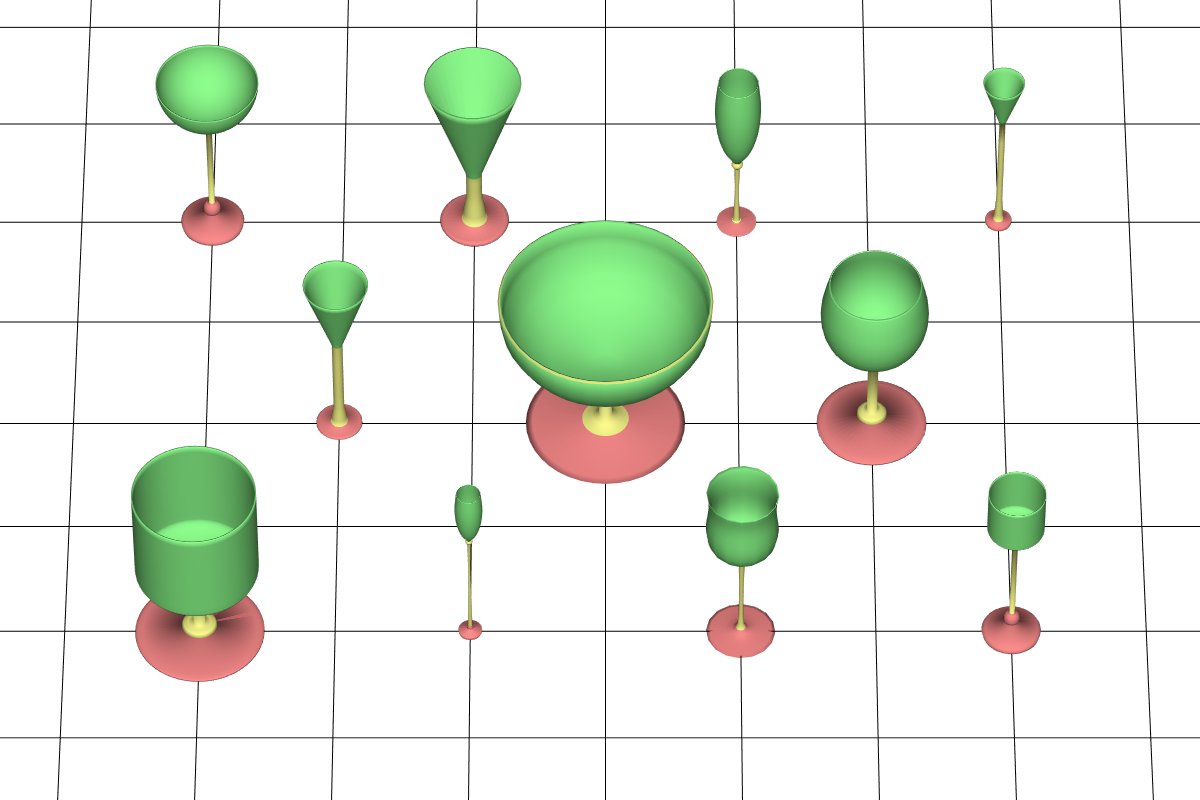}
			\caption{cosegGoblets}
		\end{subfigure}
	\end{center}
	\caption{Visualisation of some results of our 1D \ac{CNN} technique on the \ac{PSB} and Coseg datasets, with an accuracy of over 95\%.}
	\label{fig:results}
\end{figure*}

\paragraph*{Training}
First, each feature vector $\bm{X}^k$ is separately passed through a \ac{conv} layer to extract some low-level features. Sixteen 15 $\times$ 1 filters are used to produce sixteen new feature vectors, and padding is used to ensure the vectors remain a constant length. Once the output is normalised and passed through a leaky \ac{ReLU} layer, it is then max-pooled to reduce the size in half (400 components, 16 channels). This process is repeated with a \ac{conv} layer with thirty-two 11 $\times$ 1 filters. After the final pooling stage, each of the three branches provides thirty-two 200 component feature vectors.

The three branches are then merged, via a depth concatenation layer, producing ninety-six 200 component feature vectors. These are then passed through a fully connected layer with 172 neurons, and then through a dropout layer with a 50\% dropout. Finally, a fully connected layer with the number of neurons equal to the number of classes in the set is used, with a softmax activation function. The output of the softmax layer can then be used to compute the loss (we use categorical cross entropy) in order to back-propagate through the network to update parameters.

Similar to a \ac{NN}, there are two learning passes. The feed-forward pass produce a label prediction, and the back-propagation updates parameters to reduce the prediction error. These passes are repeated for a set number of iterations (set to 50, using a learning rate in the log-space between -2 and -4 and a momentum of 0.9). The label probabilities that are produced after training are subsequently used for graph-cut post-refinement. Results are reported in \textbf{1D CNN} column, Table \ref{table:results}.

\subsection{Graph-Cut Refinement}
\label{sec:refinement}
A trained model (\ac{NN}, \ac{RF}, \ac{CNN}) can predict a label for a face with a set of probabilities. The probability indicates how likely a face belongs to a particular class. However, inconsistencies of predicted labels can arise between adjacent faces on the mesh because the classification does not take face adjacency into account. This causes incorrect segmentations and reduced accuracies. Here, we utilise the multi-label alpha-expansion graph-cut technique \cite{boykov2001} to refine the segmentation results.

Let $u, v \in T$ be two faces in a mesh, where $T$ is the set of all faces. Let $N_u$ be the set of neighbouring faces of $u$. We can optimise the labels of all $u \in T$ by solving:
$$
\min_{l_u, u \in T} \sum_{u \in T} \xi_D (u, l_u) + \lambda \sum_{u \in T, v \in N_u} \xi_S (u, v, l_u, l_v, f_u, f_v)
$$
where $\lambda$ is a non-negative constant used to balance the influence of the two terms and $\xi_D (u, l_u) = -\log(\bm{p}_u(l_u))$ is a data term that penalises low probability of assigning a label $l_u$. The second term $\xi_S$ incurs a large penalty when the dihedral angle between two adjacent faces is small (i.e. the faces cause a concavity) or the distance between two features is high, and is given by:
\[
\xi_S (u, v, l_u, l_v) = \begin{cases}
0, & \text{if $l_u = l_v$} \\
-\log(\theta_{uv} / \pi) \varphi_{uv}, & \text{otherwise}
\end{cases} 
\]
where the cost is based on the dihedral angle ($\theta_{uv}$) and edge length ($\varphi_{uv}$) between faces  $u$ and $v$. This formulation has been applied commonly in \cite{guo2015, sidi2011,meng2013}.

Here, we propose to modify $\xi_S$:
\[
\xi_S (u, v, l_u, l_v, f_u, f_v) = -\log(\theta_{uv} / \pi) - \omega|| f_u - f_v ||_2
\]
by replacing the distance term  $\varphi_{uv}$ with a geometric feature term $\omega|| f_u - f_v ||_2$. It promotes similar classification label if the Euclidean distance between features $f_u$, $f_v$ of face $u$ and $v$ is small. A constant ($\omega$), is used to balance the weight of the concavity and feature terms.
We use \ac{AGD} (Section~\ref{sec:features}) as the feature $f$ as it helps to smooth out inconsistent labels, and improves the refinement accuracy (Section~\ref{sec:results}).

\begin{figure*}[t]
	\begin{center}
		\begin{subfigure}[b]{0.32\linewidth}
			\includegraphics[width=\textwidth,height=4.5cm]{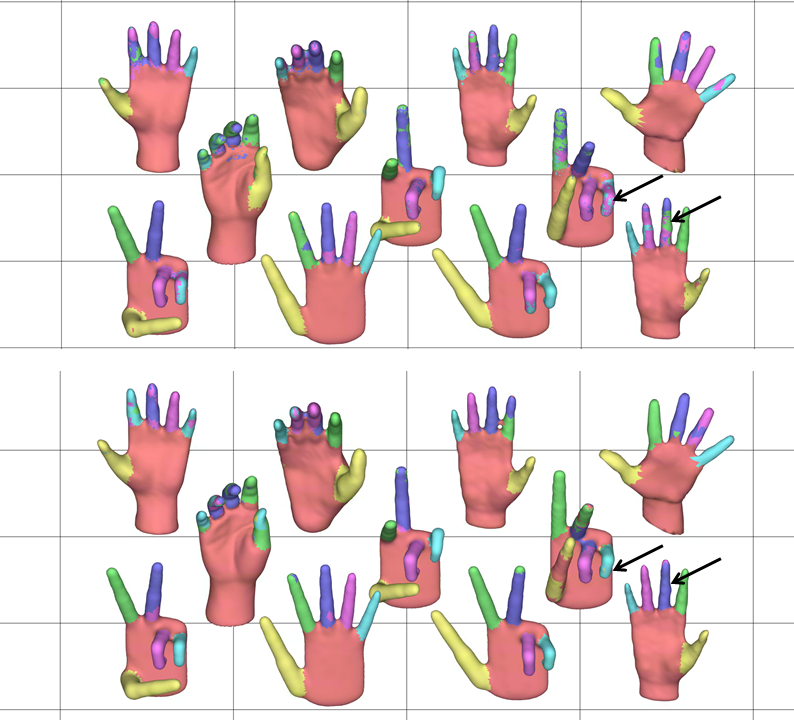}
			\caption{Hand}
		\end{subfigure}
		\begin{subfigure}[b]{0.32\linewidth}
			\includegraphics[width=\textwidth,height=4.5cm]{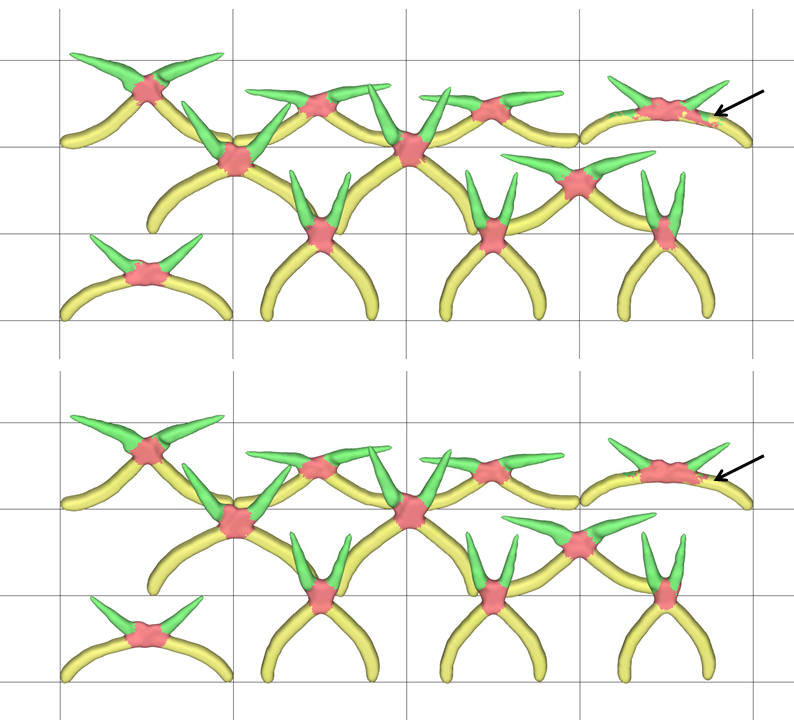}
			\caption{Plier}
		\end{subfigure}
		\begin{subfigure}[b]{0.32\linewidth}
			\includegraphics[width=\textwidth,height=4.5cm]{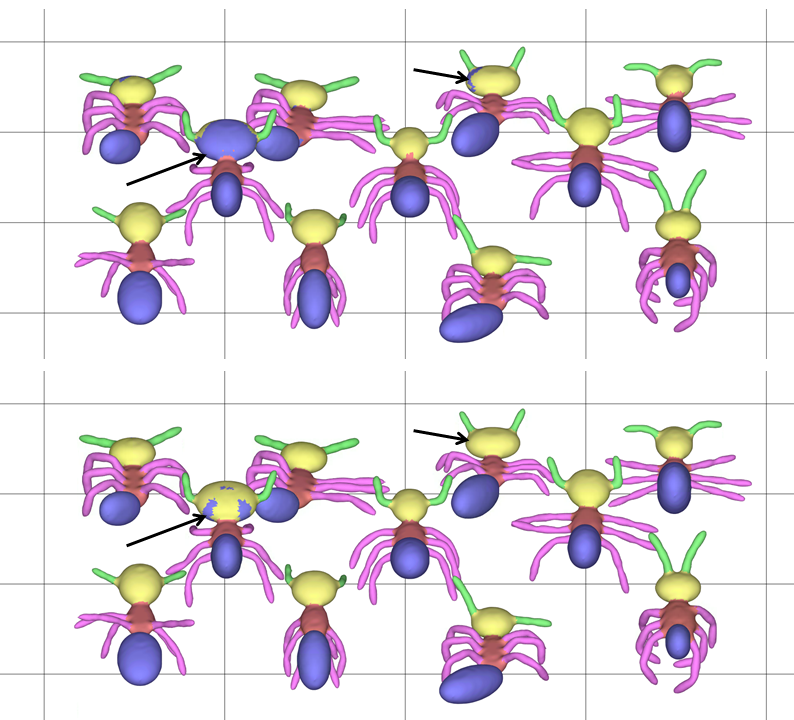}
			\caption{Ant}
		\end{subfigure}
	\end{center}
	\caption{Visual comparison of our 1D \ac{CNN} (bottom row) and TOG15 \cite{guo2015} (top row). Our method performs better on certain meshes (see arrows on the figures)}
	\label{fig:comparisonresults}
\end{figure*}


\section{Experiments and Results}
\label{sec:results}

\begin{figure*}[t]
	\begin{center}
	\begin{subfigure}[b]{0.32\linewidth}
		\includegraphics[width=\textwidth]{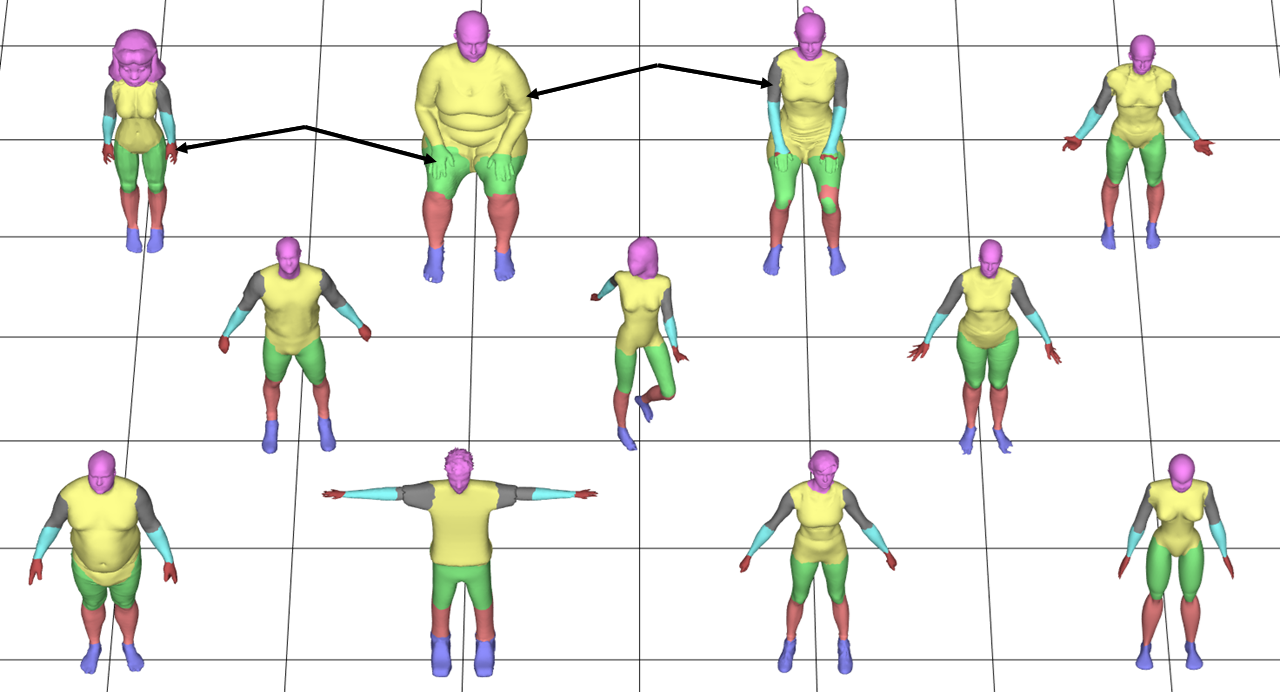}
	\end{subfigure}
	\begin{subfigure}[b]{0.32\linewidth}
		\includegraphics[width=\textwidth]{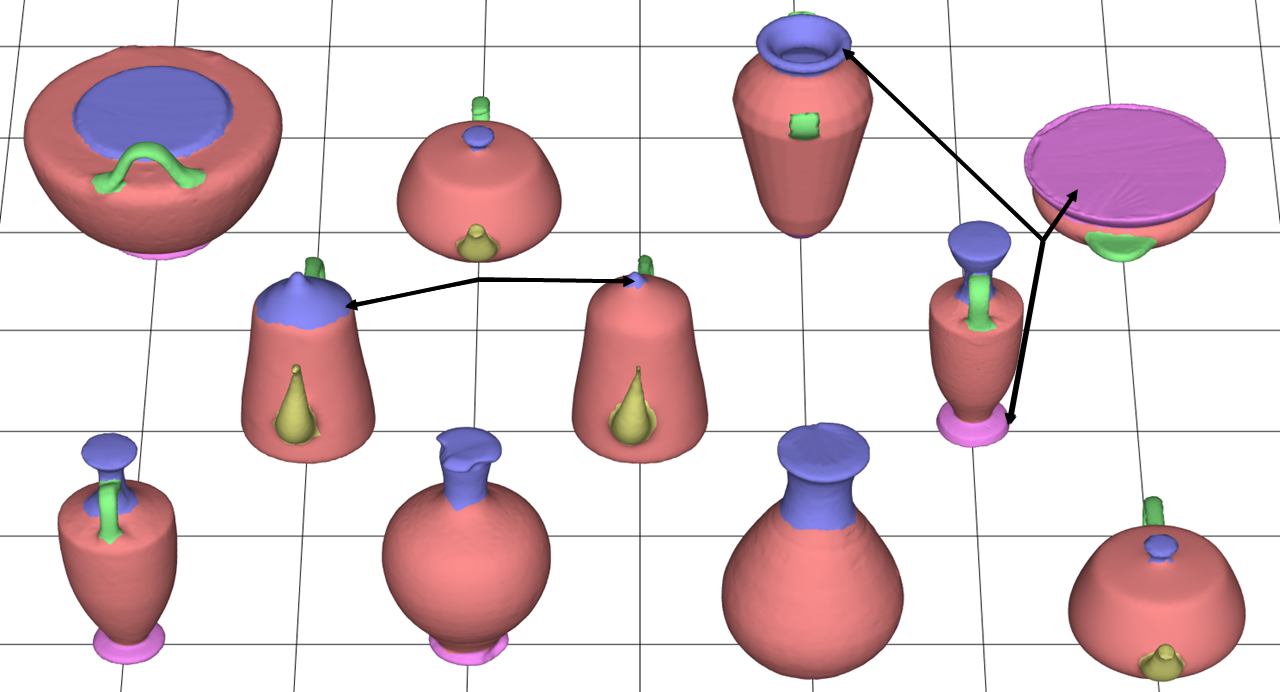}
	\end{subfigure}
	\begin{subfigure}[b]{0.32\linewidth}
		\includegraphics[width=\textwidth]{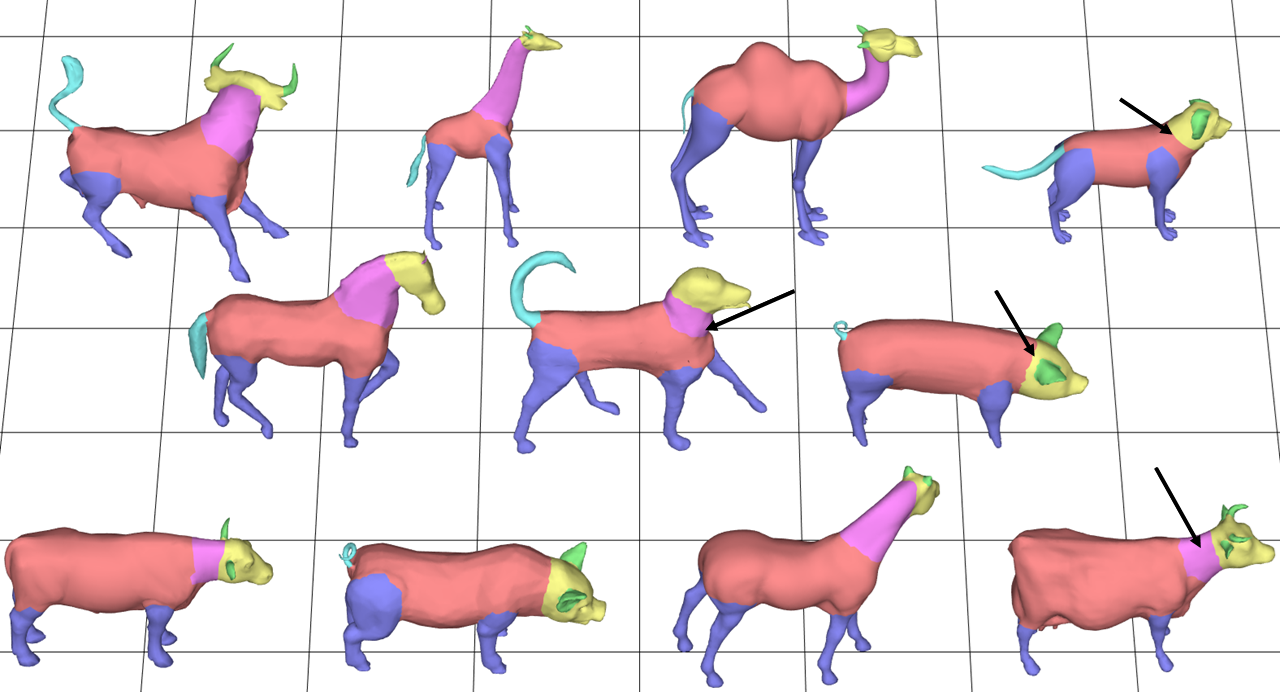}
	\end{subfigure}
	
	\begin{subfigure}[b]{0.32\linewidth}
		\includegraphics[width=\textwidth]{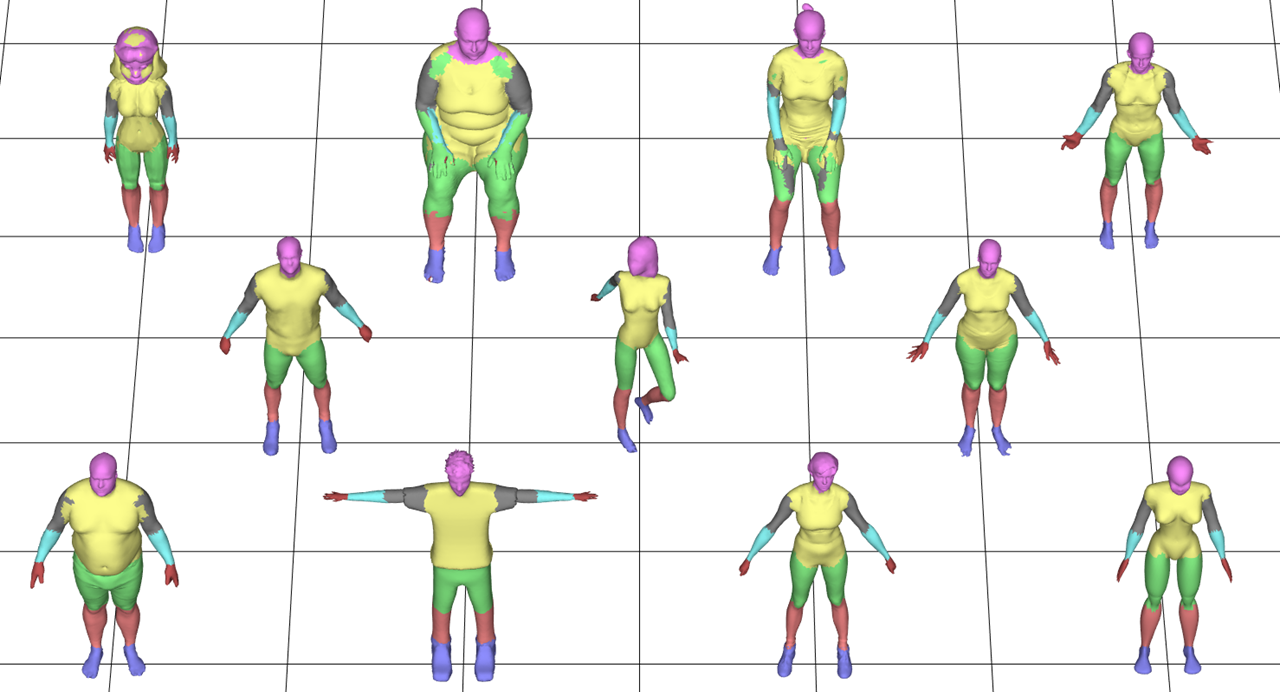}
		\caption{Human}
	\end{subfigure}
	\begin{subfigure}[b]{0.32\linewidth}
		\includegraphics[width=\textwidth]{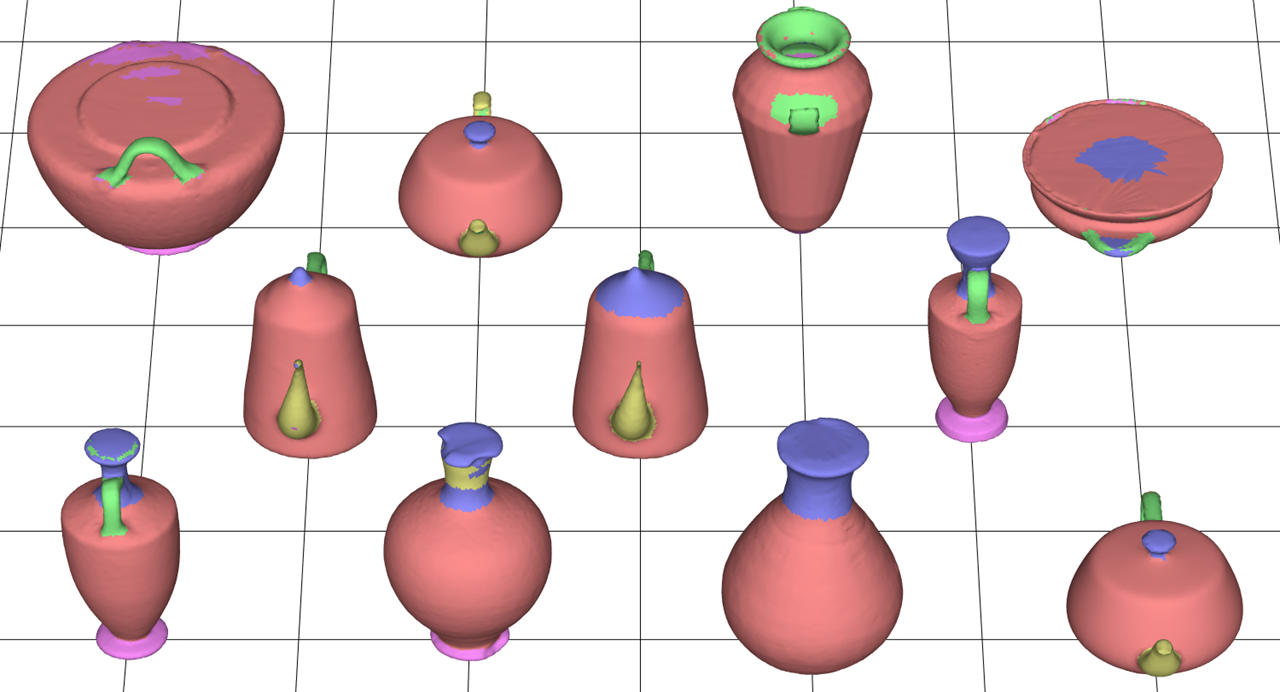}
		\caption{Vase}
	\end{subfigure}
	\begin{subfigure}[b]{0.32\linewidth}
		\includegraphics[width=\textwidth]{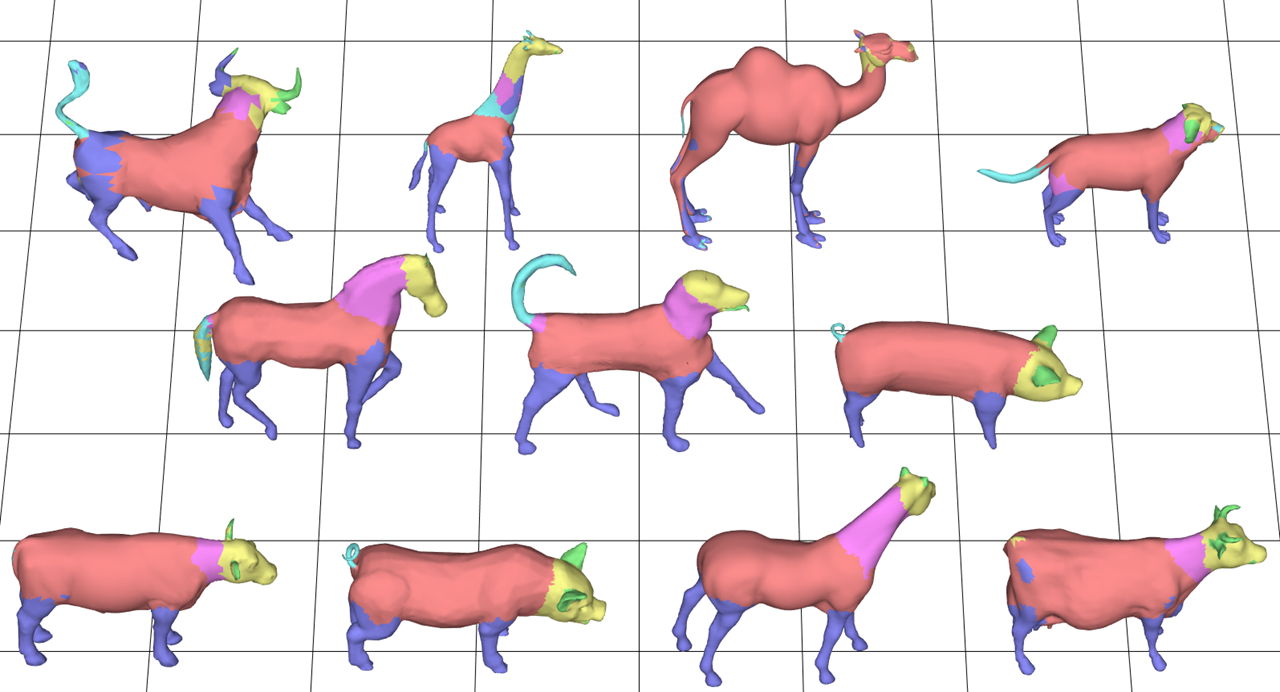}
		\caption{Fourleg}
	\end{subfigure}
\end{center}
	\caption{Visualisation of sets where our method achieved sub 90\% accuracy. Top row shows ground truth, bottom row shows our 1D \ac{CNN} results. (a) shows where poor ground truth (arrows) resulted in loss in accuracy. (b) shows where inconsistent ground truths and large variations (arrows) resulted in poor results. (c) shows where outliers in the set (back row left 3) and inconsistencies in the ground truths (arrows) caused a loss in accuracy}
	\label{fig:badresults}
\end{figure*}
In this section, we outline the experiments and accuracy measure used to evaluate each deep learning technique on mesh segmentation. Then we discuss the results and put forward our observations. 

All experiments were conducted on the \ac{PSB} \cite{chen2009} and Coseg \cite{sidi2011} datasets, which are widely used datasets for evaluating mesh segmentation techniques \cite{kalogerakis2010,guo2015}. The \ac{PSB} dataset contains 19 sets with 20 meshes per set. Similar to \cite{shu2016}, we omit three sets (Bust, Bearing and Mech) from our results in Table~\ref{table:results}, because these sets are either inconsistently labelled or contain meshes with too much variance within the set (Further discussion is provided at the end of this section). The Coseg dataset has 8 sets with between 12 and 44 meshes per set. The dataset also includes 3 large sets with 200 to 400 meshes per set. We ran 3 different types of experiments:
\begin{itemize}[leftmargin=*]
	\vspace{-5pt}
	\setlength\itemsep{-2pt}
	\item Leave-one-out cross validation - a single mesh is removed from the training set and used exclusively for testing. This is repeated for all meshes in the set.
	\item 5-fold cross validation - the set is split into 5 equal (or close to equal) subsets. In each run, a single subset is left out of the training set and used for testing. This is repeated for all 5 folds. The training \& testing splits used will be publicly available.
	\item Fixed training/testing split - we run experiment using the training/testing splits defined in \cite{evangelos2017} for each set.
\end{itemize}
The \ac{PSB} dataset was used in all experiments. The Coseg dataset was only used in the 5-fold and fixed training/testing split experiments \cite{evangelos2017} as running leave-one-out cross validation is a lengthy process.

For all experiments we use the accuracy measure: 
\[
Accuracy(l,gt) = \frac{\sum_{t \in T} a_t  \delta (l_t = gt_i)}{\sum_{t \in T} a_i}
\]
where $a_t$, $l_t$ and $gt_t$ are respectively the area, the predicted label and the ground truth label of triangle $t$. $\delta (l_i = gt_i)$ is assigned to 1, if the predicted label is the same as the ground truth; otherwise 0. This is similar to \cite{kalogerakis2010,guo2015,sidi2011}

Finally, as pointed out by Kalogerakis et al \cite{evangelos2017}, there is no publicly available implementation for Guo et al's architecture \cite{guo2015}. Similarly, we use our own faithful reimplementation of Guo et al's architecture, closely following the details in the paper. We  tried Matlab (MatConvNet) and Python (TensorFlow) implementations, and both show similar results. We reported Python's results as they are marginally better. Both reimplementations are internally validated by three independent researchers from two research teams at Swansea University. Both source codes are available for external validation. Though the reported accuracy of \cite{guo2015} is lower than that reported in the original paper, the reproduced accuracy is still higher than the results reported in \cite{evangelos2017}. We believe that our reimplementation is faithful.

\begin{table}
	\newcommand{\tbf}{\textbf}
	\newcommand{\tul}{\underline}
	{\resizebox{\columnwidth}{!}{                                                  
		\begin{tabular}{@{}L{1.9cm} @{}C{1cm} C{1cm} C{1cm} C{1.2cm} C{1.1cm}@{}}           
			\toprule                                                                    
								& \tbf{PCA}		& \tbf{AE} 			& \tbf{AE} 		& \tbf{ToG15} 			& \tbf{1D} 			\\
								& \tbf{\& NN} 	& \tbf{\& RF} 		& \tbf{\& NN} 	& \tbf{\cite{guo2015}} 	& \tbf{CNN} 		\\ \midrule
			\tbf{Airplane} 		& 92.97       	& 92.62             & 92.53       	& 94.56                	& \tbf{96.52}		\\ \midrule
			\tbf{Ant}      		& 95.15       	& 95.17             & 95.15       	& 97.55                	& \tbf{98.75} 		\\ \midrule
			\tbf{Armadillo}		& 88.21       	& 88.43             & 87.79       	& 90.90                	& \tbf{93.74} 		\\ \midrule
			\tbf{Bird}     		& 85.14       	& 88.93       		& 88.20       	& 86.20                	& \tbf{91.67}   	\\ \midrule
			\tbf{Chair}    		& 95.55       	& 95.69             & 95.61      	& 97.07       			& \tbf{98.41}   	\\ \midrule
			\tbf{Cup}      		& 95.09       	& 97.95             & 97.82       	& 98.95				    & \tbf{99.73}   	\\ \midrule
			\tbf{Fish}    		& 94.41      	& 96.21				& 95.31       	& 96.16          		& \tbf{96.44} 		\\ \midrule
			\tbf{Fourleg}  		& 83.61       	& 83.99             & 82.32      	& 81.91                	& \tbf{86.74} 		\\ \midrule
			\tbf{Glasses}   	& 94.22       	& 96.57       		& 96.42 		& 96.95					& \tbf{97.09}		\\ \midrule
			\tbf{Hand}      	& 78.33       	& 73.76             & 70.49       	& 82.47                	& \tbf{89.81} 		\\ \midrule
			\tbf{Human}  		& 87.03       	& 86.69             & 81.45       	& 88.90			        & \tbf{89.81}		\\ \midrule
			\tbf{Octopus}   	& 96.93       	& 96.99             & 96.52       	& 98.50			 		& \tbf{98.63}		\\ \midrule
			\tbf{Plier}     	& 93.75       	& 92.59             & 91.53       	& 94.54                	& \tbf{95.61} 		\\ \midrule
			\tbf{Table}     	& 99.22		 	& 99.18 		    & 99.17 		& 99.29					& \tbf{99.55}		\\ \midrule
			\tbf{Teddy}    		& 98.07 		& 98.24				& 98.20 		& 98.18          		& \tbf{98.49}		\\ \midrule
			\tbf{Vase}     		& 79.73       	& 82.07             & 80.24       	& 82.81			      	& \tbf{85.75}   	\\ \midrule
			\midrule
			\tbf{Average}   	& 91.09       	& 91.57             & 90.61       	& 92.79  				& \tbf{94.80} 		\\ \midrule
		\end{tabular}}}                                                                   
	\captionof{table}{Experimental results for leave-one-out cross validation on the \ac{PSB} dataset \cite{chen2009}. \textbf{Bold}: highest accuracy.}
	\label{table:results}
	\vspace{-0.7cm}
\end{table}

\begin{figure*}
	\begin{center}
		\begin{subfigure}[b]{0.32\linewidth}
			\includegraphics[width=\textwidth]{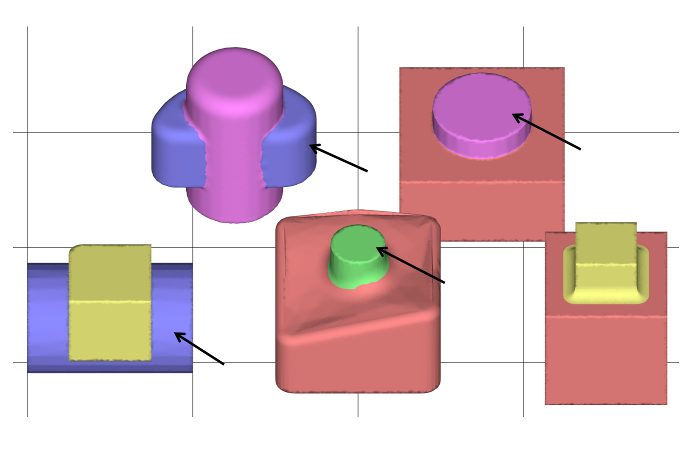}
			\caption{Mech}
		\end{subfigure}
		\begin{subfigure}[b]{0.32\linewidth}
			\includegraphics[width=\textwidth]{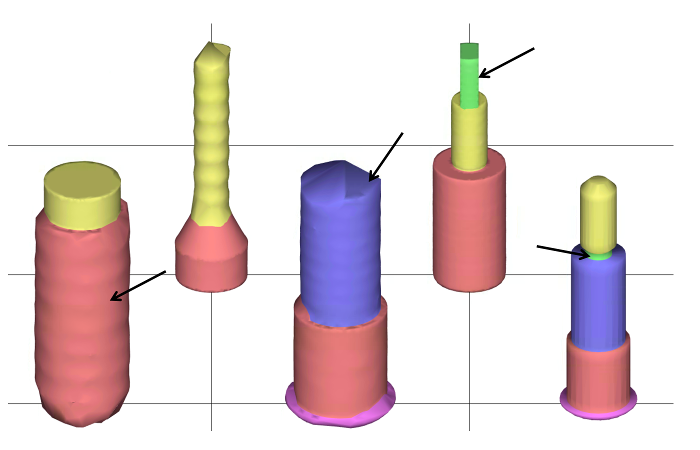}
			\caption{Bearing}
		\end{subfigure}
		\begin{subfigure}[b]{0.32\linewidth}
			\includegraphics[width=\textwidth]{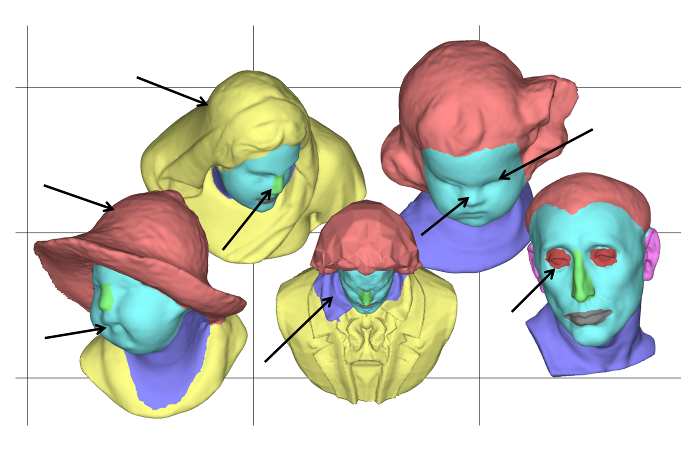}
			\caption{Bust}
		\end{subfigure}
	\end{center}
	\caption{\textbf{Ground truth} examples from the 3 omitted sets (Mech, Bearing, Bust). Label inconsistencies can be seen throughout. Mech (a) shows segment inconsistencies where cylindrical shapes are labelled both purple and green (front row, centre and back row). Also, the blue segments shown on the two meshes are the only blue segments in the set, and are both topologically dissimilar. Bearing (b) shows segment inconsistencies where similar shaped regions (threaded parts) have several different labels. Bust (c) shows poor segment boundaries where the neck extends on to the clothing (front row, centre). Additionally, it contains inconsistent segments where the hats and hair are one segment but back left has a clothing segment over the top of the head. Finally, a few labels are missing throughout. For example, not all lips, noses and eyes are properly and consistent labelled. Some models are missing some of these segments and others are missing them all from the ground truth (e.g. nose of back right, eyes of 4 of the shown models, lips of front left and back right). Arrows show examples of badly or inconsistent ground truth labelling.}
	\label{fig:badgt}
\end{figure*}

\paragraph*{\textbf{Leave-one-out Cross Validation}}
Experimental results for Leave-one-out cross validation are shown in Table~\ref{table:results}, where the columns \textbf{PCA \& NN}, \textbf{AE \& RF}, \textbf{AE \& NN} and \textbf{1D CNN} correspond to the techniques discussed in Sections \ref{sec:neuralnet}-\ref{sec:cnn} and column \textbf{TOG15} shows the results using \cite{guo2015}. Some visual results of our 1D \ac{CNN} technique are shown in Figure~\ref{fig:results}.

A direct comparison of our 1D \ac{CNN} and \cite{guo2015} shows that our method achieves higher accuracies on all of the sets. The majority of the sets see a large improvement in accuracy, especially some of the harder sets (Armadillo, Hand, Vase). Five sets (Fish, Glasses, Octopus, Table, Teddy), which have very high accuracies ($> 96\%$) in \cite{guo2015}, also show slight improvement.
Figure~\ref{fig:results} and \ref{fig:comparisonresults} show some visual comparisons of the results.

We further investigate the sets with poorer results ($<$ 90\%) and observe several problems.
\begin{itemize}[leftmargin=*]
\vspace{-5pt}
\setlength\itemsep{-2pt}
\item First, the Human set (Figure~\ref{fig:badresults}, column (a)) is badly and non-consistently labelled in general, and there is insufficient support to train a proper model (see arrows). This is challenging for any machine learning technique, making all techniques fail to achieve high accuracies ($>$ 90\%).
\item Second, the Vase set (Figure~\ref{fig:badresults}, column (b)) contains meshes that are significantly different from the rest. As indicated by arrows, there is a mesh (back row, second from right) that contains a segment usually defined as the base of a vase (purple segment), in place of the part which is typically on the top of the vase (blue segment in other meshes). Also there are two meshes that are almost identical (middle row), but the label is very different (e.g. blue segment).
\item Third, the Fourleg set (Figure~\ref{fig:badresults}, column (c)) contains three meshes that are very different from all other meshes in the set (back row, left most 3). Label inconsistencies are also present where the majority of meshes contain a neck segment (purple), but some meshes do not (see arrows).
\item Finally, each of the fingers in the Hand set (Figure~\ref{fig:comparisonresults}) are considered separate segments in the ground truth. It is very hard to achieve good results using features alone for such set. We believe the result can be improved with a correspondence matching technique.
\end{itemize}

For completeness, the results for the sets we omitted from Table~\ref{table:results} are as follows. Accuracies of 88.67\%, 70.06\% and 88.53\% \cite{guo2015}, and 89.69\%, 61.97\% and 88.14\% (our 1D \ac{CNN}) were achieved for the Bearing, Bust and Mech sets respectively. Following \cite{shu2016}, these sets were omitted due to ground truth inconsistencies (Figure \ref{fig:badgt}) and lack of sufficient training data. These are reflected in the lower accuracies of both methods.

Next we analyse the performance of our other deep learning techniques, \ac{PCA} \& \ac{NN}, \ac{AE} \& \ac{RF}, and \ac{AE} \& \ac{NN}. We note that, although they perform worse than \ac{CNN} techniques, in general, their performance is only marginally worse. In sets that have consistent and well-defined labels (Fish and Teddy), \ac{AE} \& \ac{RF} performs better than the 2D \ac{CNN} technique. This is interesting as these \ac{NN} models consists of 2-3 layers, and require much shorter training time than \ac{CNN} techniques.

Finally we compare the use of two different classifiers \ac{AE} \& \ac{RF} and \ac{AE} \& \ac{NN} on the same set of encoded features. As shown in Table~\ref{table:results}, the results from using an \ac{RF} classifier are almost exclusively better than using only the \ac{NN} model alone. This may be explained by the fact that both the \ac{AE} network and the \ac{RF} classifier are two different techniques, and are separately trained. It suggests that there is complementary improvement overall.

Our conclusion is that, compared to \cite{guo2015}, if there is a sufficiently large number of good meshes with consistent labels across the set, the new features and 1D \ac{CNN} architecture can improve performance. Our technique also does not require parameter tuning for features reshaping or sampling to 2D images.

\paragraph*{\textbf{5 Fold Cross Validation}}
\begin{table}
	\newcommand{\tbf}{\textbf}
	\newcommand{\tul}{\underline}
	\newcommand{\mr}{\multirow{2}{*}}
	{\resizebox{\columnwidth}{!}{                                                    
			\begin{tabular}{@{}L{1.6cm} C{1cm} @{}C{1cm} @{}C{1cm} @{}C{1cm} @{}C{1cm} @{}C{1cm} @{}C{1.1cm}@{}}          
				\toprule                  
								& \multicolumn{6}{c}{\tbf{1D CNN}} 																& \tbf{ToG15}				\\
								& \mr{\tbf{1B}} & \mr{\tbf{2B}}	& \mr{\tbf{3B}}	& \mr{\tbf{4B}} & \tbf{1B}		& \tbf{3B}		& \mr{\tbf{\cite{guo2015}}} \\ 
								& 				&				&				&				& \tbf{(600)}	& \tbf{(600)}	&							\\ \midrule
				\tbf{Airplane} 	& 94.93     	& 95.57     	& 95.65     	& 95.60    		& 94.24			& 94.31			& 93.43 					\\ \midrule
				\tbf{Ant}      	& 98.24     	& 98.75     	& 98.82     	& 98.75    		& 97.15			& 97.23 		& 96.91						\\ \midrule
				\tbf{Armadillo}	& 93.08     	& 93.29     	& 93.47     	& 93.56    		& 91.02 		& 92.17 		& 87.05						\\ \midrule
				\tbf{Bird}     	& 90.86     	& 91.14     	& 91.34     	& 91.82    		& 90.69			& 90.80   		& 90.00						\\ \midrule
				\tbf{Chair}    	& 97.72     	& 98.03     	& 98.14     	& 98.39    		& 96.57			& 97.10   		& 96.43						\\ \midrule
				\tbf{Cup}      	& 99.62     	& 99.65     	& 99.69     	& 99.65			& 99.45			& 99.65   		& 99.13						\\ \midrule
				\tbf{Fish}    	& 96.47     	& 96.69			& 96.75     	& 96.82			& 96.39	 		& 96.68		 	& 95.99						\\ \midrule
				\tbf{Fourleg}  	& 87.10     	& 87.78     	& 88.23     	& 88.43    		& 86.38			& 87.50 		& 84.92						\\ \midrule
				\tbf{Glasses}   & 96.68     	& 96.72			& 96.89			& 96.94			& 96.25			& 96.61			& 96.31						\\ \midrule
				\tbf{Hand}      & 88.86     	& 89.33     	& 89.66     	& 89.64    		& 87.40			& 88.45 		& 80.31						\\ \midrule
				\tbf{Human}  	& 87.50     	& 88.81     	& 89.02     	& 88.85			& 87.34			& 88.49			& 82.51						\\ \midrule
				\tbf{Octopus}   & 98.54     	& 98.67     	& 98.71     	& 98.75			& 98.51			& 98.61			& 98.39						\\ \midrule
				\tbf{Plier}     & 95.41     	& 95.36     	& 95.52     	& 95.59    		& 95.29			& 95.34 		& 95.23						\\ \midrule
				\tbf{Table}     & 99.61			& 99.62			& 99.62			& 99.62			& 99.59			& 99.62			& 99.08						\\ \midrule
				\tbf{Teddy}    	& 98.39			& 98.37			& 98.35			& 98.40			& 96.67			& 97.06		 	& 95.90						\\ \midrule
				\tbf{Vase}     	& 84.43     	& 86.35     	& 87.10     	& 86.11			&  84.06 		& 84.66   		& 81.08						\\ \midrule
				\midrule                    	            	            	           					
				\tbf{Average}   & 94.22     	& 94.63     	& \tbf{94.81}     	& \tbf{94.81}  		& 	93.57	 		& 94.02		 	& 92.04						\\ \bottomrule
	\end{tabular}}}
	\captionof{table}{5-fold cross validation labeling accuracies for the \ac{PSB} dataset \cite{chen2009}. Results of our 1D \ac{CNN} with differing number of branches are shown (\tbf{1B}, \tbf{2B}, \tbf{3B}, \tbf{4B}), as well as using the same features as ToG15 \cite{guo2015} (\tbf{1B (600)}, \tbf{3B (600)})}
	\label{table:results-psb}
	\vspace{-0.8cm}
\end{table}

The experimental results in Table~\ref{table:results-psb} show a comparison of running our 1D \ac{CNN} architecture with different numbers of branches on the \ac{PSB} dataset. As shown in the table (Columns \textbf{1B}, \textbf{2B}, \textbf{3B}, \textbf{4B}), performance steadily increases as the number of branches increase, up until it plateaus at 3-4 branches.  A direct comparison to the result of 2D \ac{CNN} (\textbf{ToG15 \cite{guo2015}}) shows that our method outperforms the 2D architecture for all sets, even using a single branch network. It also supports empirically our choice of using a 3-branch network as it is faster to train whilst reaching similar performance as compared to that of 4-branch.

Table~\ref{table:results-psb} also shows that, when using the same set of features in \cite{guo2015}, whether it is one (Column \textbf{1B (600)}) or three branches (Column \textbf{3B (600)}), our architecture can still outperform, with the latter giving better results. These results show that the use of 1D data and filters is useful, and that the multi-branch architecture (with multi-scale features) and the addition of new features (including our proposed more robust conformal factors) all separately contribute to the improvement.

We further show experimental results using the Coseg dataset \cite{sidi2011} in Table~\ref{table:results-coseg} and Figure~\ref{fig:results}. We see a large improvement over the 2D \ac{CNN} \cite{guo2015} for the smaller datasets (over 4\% on average, Table~\ref{table:results-coseg} left), and notice a larger improvement when the datasets have hundreds of meshes (over 6\% on average, Table~\ref{table:results-coseg} right). This shows that our model can generalize well when sufficient training data is provided, even if there are large variations in the meshes in the sets (e.g. VasesLarge, AliensLarge). This supports our earlier conclusion that, given a large set of well labelled meshes, our method can be effectively trained for good performance, and can handle largely varying meshes in the set.

\paragraph*{\textbf{Fixed training/testing splits}}
A final set of experiments use the training/testing splits defined in the concurrent work \cite{evangelos2017}, and use the \ac{PSB} and Coseg datasets. Table~\ref{table:results-1fold} shows the results of these experiments for 2D \ac{CNN} \cite{guo2015}, Projective \ac{CNN} \cite{evangelos2017}, and our 1D \ac{CNN}. (The results for 2D \ac{CNN} \cite{guo2015} and Projective \ac{CNN} \cite{evangelos2017} are copied from \cite{evangelos2017} for direct comparison). It shows that our 1D \ac{CNN} technique clearly outperforms the existing 2D \ac{CNN} architecture \cite{guo2015}, and also performs comparably with the concurrent work Projective \ac{CNN} architecture \cite{evangelos2017}. Note however that these experiments do not fully evaluate the method for each set because not all meshes are used at least once for testing. We include these results for completeness.

\begin{table}
\begin{tabular}{@{}C{\columnwidth}@{}}

	\begin{minipage}{\columnwidth}
		\newcommand{\tbf}{\textbf}
		\newcommand{\tul}{\underline}
		{\begin{center}
		\vspace{-0.7cm}
		\begin{tabular}{@{}L{0.5\columnwidth}@{}  @{}L{0.5\columnwidth}@{}}
		{\begin{center}
			\vspace{-0.1cm}
				\resizebox{0.48\columnwidth}{!}{                                                   
					\begin{tabular}{@{}L{1.8cm} @{}C{1.4cm} @{}C{1.6cm}@{}}          
						\toprule                  
						\textit{\tbf{SmallSet}} & \tbf{1D CNN}	& \tbf{ToG15\cite{guo2015}} 		\\ \midrule
						\tbf{Candelabra}		& 93.58     	& 91.55 			\\ \midrule
						\tbf{Chairs}   			& 97.75     	& 93.48			\\ \midrule
						\tbf{Fourleg}			& 94.12     	& 90.75			\\ \midrule
						\tbf{Goblets}  			& 97.80     	& 92.79		\\ \midrule
						\tbf{Guitars}  			& 98.03     	& 97.04			\\ \midrule
						\tbf{Irons}    			& 89.89     	& 80.90			\\ \midrule
						\tbf{Lamps}    			& 86.74     	& 81.52		\\ \midrule
						\tbf{Vases}  			& 92.47     	& 89.42			\\ \midrule \midrule
						\tbf{Average}			& \tbf{93.80}			& 89.68		\\ \midrule
				\end{tabular}}
		\end{center}}                                                                   
		&
		{\begin{center}
			\vspace{-2.4cm}
				\resizebox{0.45\columnwidth}{!}{                                                     
					\begin{tabular}{@{}L{1.4cm} @{}C{1.4cm} @{}C{1.6cm}@{}}           
						\toprule                  
						\textit{\tbf{LargeSet}} & \tbf{1D CNN}	& \tbf{ToG15\cite{guo2015}} \\ \midrule
						\tbf{Vases} 		& 95.88     	& 87.57						\\ \midrule 
						\tbf{Chairs}  		& 97.71     	& 92.68						\\ \midrule
						\tbf{Aliens}  		& 97.84     	& 91.93						\\ \midrule  \midrule
						\tbf{Average}	& \tbf{97.14}			& 90.73						\\ \midrule
				\end{tabular}}
		\end{center}}                                                                  
		\end{tabular}
		\end{center}\vspace{-0.5cm}} 
		\captionof{table}{5-fold cross validation labelling accuracies for the Coseg dataset \cite{sidi2011}.}
		\label{table:results-coseg}

	\end{minipage}

\\

\begin{minipage}{\columnwidth}
		\newcommand{\tbf}{\textbf}
		\newcommand{\tul}{\underline}
		\vspace{0.3cm}
		{\resizebox{\columnwidth}{!}{                                                     	
			\begin{tabular}{L{2.6cm} C{1.2cm} C{0.8cm} C{1.3cm} C{1.9cm}@{}}          	
				\toprule                  
							& \tbf{Testing}	& \tbf{1D}	& \tbf{ToG15}			& \tbf{ShapePFCN}	\\
							& \tbf{Meshes}	& \tbf{CNN}	& \tbf{\cite{guo2015}}	& \tbf{\cite{evangelos2017}} \\ \midrule
			\tbf{psbAirplane} 		& 8     		& 95.92     & 91.60					& 93.00 		\\ \midrule
			\tbf{psbAnt}      		& 8     		& 98.72     & 97.60 				& 98.60			\\ \midrule
			\tbf{psbArmadillo}		& 8     		& 93.31     & 85.00 				& 92.80			\\ \midrule
			\tbf{psbBird}     		& 8     		& 91.04     & 83.10   				& 92.30			\\ \midrule
			\tbf{psbChair}    		& 8     		& 97.67     & 96.70   				& 98.50			\\ \midrule
			\tbf{psbCup}      		& 8     		& 94.45     & 92.10   				& 93.80			\\ \midrule
			\tbf{psbFish}    		& 8     		& 96.48		& 94.50		 			& 96.00		\\ \midrule
			\tbf{psbFourleg}  		& 8     		& 87.71     & 82.40 				& 85.00			\\ \midrule
			\tbf{psbGlasses}   		& 8     		& 96.31		& 95.30					& 96.60		\\ \midrule
			\tbf{psbHand}      		& 8     		& 91.70     & 73.80 				& 84.80			\\ \midrule
			\tbf{psbHuman}  		& 8     		& 90.58     & 85.60					& 94.50		\\ \midrule
			\tbf{psbOctopus}		& 8     		& 98.48     & 97.40					& 98.30		\\ \midrule
			\tbf{psbPlier}     		& 8     		& 95.81     & 95.20 				& 95.50			\\ \midrule
			\tbf{psbTable}     		& 8				& 99.57		& 98.50					& 99.50	\\ \midrule
			\tbf{psbTeddy}    		& 8				& 88.27		& 97.30			 		& 97.70	\\ \midrule
			\tbf{psbVase}     		& 8     		& 81.94     & 77.80   				& 86.80			\\ \midrule
			\tbf{psbAverage}   		& -     		& 93.62     & 90.24   				& 93.98		\\ \midrule \midrule
			\tbf{cosegCandelabra}   & 16     		& 94.39     & 85.90 				& 95.40			\\ \midrule
			\tbf{cosegChairs}  		& 8     		& 96.02     & 93.80					& 96.10		\\ \midrule
			\tbf{cosegFourleg}		& 8     		& 93.64     & 88.20					& 90.40		\\ \midrule
			\tbf{cosegGoblets}     	& 6     		& 99.46     & 86.10 				& 97.20			\\ \midrule
			\tbf{cosegGuitars}     	& 35			& 98.43		& 97.70					& 98.00		\\ \midrule
			\tbf{cosegIrons}    	& 6				& 84.75		& 79.70			 		& 88.00	\\ \midrule
			\tbf{cosegLamps}     	& 8     		& 84.04     & 78.00   				& 93.00			\\ \midrule
			\tbf{cosegVases}   		& 16     		& 87.55		& 84.40					& 84.80	\\ \midrule
			\tbf{cosegAverage} 		& -     		& 92.29		& 86.73					& 92.86	\\ \midrule \midrule	
			\tbf{Total Average} 	& -    			& 93.18     & 89.07		 			& 93.61	\\ \bottomrule
		\end{tabular}}}                                                                   
		\captionof{table}{Fixed training/testing split results for \ac{PSB} \cite{chen2009} and Coseg \cite{sidi2011} datasets. Training/testing splits are the same as \cite{evangelos2017}, all sets use 12 training meshes (except cosegGoblets which uses 6)}
		\label{table:results-1fold}
		\vspace{-0.6cm}	
	\end{minipage}

\end{tabular}
\end{table}

\section{Conclusion}
\label{sec:conclusion}

In this paper, we have shown a novel way of using \acp{CNN} on the geometric feature space to perform automatic mesh segmentation. Instead of casting 3D geometric features into 2D images and using 2D filters to fit an image-based \ac{CNN} pipeline, we show that the use of 1D data and filters can alleviate unnecessary inference of unrelated features. It also avoids the problem of parameter tuning for reshaping and re-sampling of features, and achieves better performance. Our novel technique clearly out-performs existing work \cite{guo2015} in terms of accuracy and can support more features and a more complex and deeper network. We have further shown a novel way of computing more consistent and robust \acl{CF} which is less sensitive to small areas of large curvature. 

We additionally performed a comprehensive and comparative study of several deep learning techniques for mesh segmentation. We showed that simpler network architectures (e.g. \acp{AE}, \acp{NN} and \acp{RF}) can still perform reasonably well using the same set of geometric features when compared to more complex \ac{CNN} models. Their training time is also significantly shorter. This suggests that if only a reasonable (not perfect) segmentation is required for downstream application, \ac{AE}, \ac{NN} and \ac{RF} would be a good choice.

We have also shown some labelling problems in the \ac{PSB} dataset which is commonly used as a segmentation benchmark. For example, there is an insufficient number of meshes in certain sets to cover a large variation of shape and topology and some of the segmentation boundaries in the ground truth labels are not well-defined or consistent (e.g. Human, Bearing, Bust).

Finally, we release the data and code of all techniques discussed in this paper, helping the research of supervised mesh segmentation in the community.

\section*{Acknowledgements}
\vspace{-0.2cm}
\noindent David George is fully-funded by a PhD Studentship (DTG) from the UK Engineering and Physical Sciences Research Council.

\section*{References}





\bibliographystyle{elsarticle-num}
\bibliography{references.bib}

\end{document}